\documentclass[12pt]{article}

\usepackage[numbers,sort&compress]{natbib}
\usepackage{graphicx}
\usepackage{amsmath}
\usepackage{amssymb}

\newcommand{\Tr}{\ensuremath{\mathop{\mathrm{Tr}}}}

\def\ni{\noindent}
\def\be{\begin{equation}}
\def\ee{\end{equation}}
\def\bea{\begin{eqnarray}}
\def\eea{\end{eqnarray}}
\def\bsp{\be\begin{split}}
\def\la{\langle}
\def\ra{\rangle}

\def\wt{\widetilde}

\def\lr{\leftrightarrow}

\def\G{\Gamma}
\def\D{\Delta}
\def\L{\Lambda}
\def\S{\Sigma}
\def\a{\alpha}
\def\b{\beta}
\def\g{\gamma}

\def\d{\delta}
\def\e{\epsilon}
\def\m{\mu}

\def\s{\sigma}
\def\r{\rho}
\def\l{\lambda}
\def\t{\tau}
\def\o{\omega}

\def\T{\theta}
\def\TO{\theta_0}

\def\vt{\vartheta}

\def\p{\partial}

\def\lr{\leftrightarrow}

\makeatletter

\newcommand{\Rmnum}[1]{\expandafter\@slowromancap\romannumeral #1@}
\makeatother

\renewcommand{\title}[1]{\vbox{\center\LARGE{#1}}\vspace{5mm}}
\renewcommand{\author}[1]{\vbox{\center\large{#1}}\vspace{5mm}}
\newcommand{\address}[1]{\vbox{\center\em#1}}
\newcommand{\email}[1]{\vbox{\center\tt#1}\vspace{5mm}}

\begin{document}

\begin{titlepage}
\hfill {\tt HU-EP-08/13}\\
\title{\vspace{1.0in} {\bf BPS Wilson Loops on $S^2$ at Higher Loops}}

\author{Donovan Young}

\address{Humboldt-Universit\"at zu Berlin, Institut f\"ur Physik,\\
    Newtonstra\ss e 15, D-12489 Berlin,
    Germany}

\email{dyoung@physik.hu-berlin.de}

\abstract{\ni We consider supersymmetric Wilson loops of the variety
constructed by Drukker, Giombi, Ricci, and Trancanelli, whose
spatial contours lie on a two-sphere. Working to second order in the
't Hooft coupling in planar ${\cal N}=4$ Supersymmetric Yang-Mills
Theory (SYM), we compute the vacuum expectation value of a
wavy-latitude and of a loop composed of two longitudes. We evaluate
the resulting integrals numerically and find that the results are
consistent with the zero-instanton sector calculation of Wilson
loops in 2-d Yang-Mills on $S^2$ performed by Bassetto and Griguolo.
We also consider the connected correlator of two distinct latitudes
to third order in the 't Hooft coupling in planar ${\cal N}=4$ SYM.
We compare the result in the limit where the latitudes become
coincident to a perturbative calculation in 2-d Yang-Mills on $S^2$
using a light-cone Wu-Mandelstam-Leibbrandt prescription. We are not
able to calculate the SYM result at the required order in the
separation between the latitudes necessary for a match with 2-d
Yang-Mills; the result, however, does not preclude such a match.}

\end{titlepage}

\section{Introduction and results}

The study of Wilson loops in ${\cal N}=4$ supersymmetric Yang-Mills
theory \cite{Maldacena:1998im,Drukker:1999zq} has provided a unique
and rich avenue for probing the AdS/CFT correspondence
\cite{Maldacena:1997re} as well as the theory itself. Certain loops
which respect some of the supersymmetries of the underlying theory
have been analyzed with great success. Loops with arbitrary shape may
be constructed with enough supersymmetry to yield trivial vacuum
expectation values \cite{Zarembo:2002an,Guralnik:2003di}, a result
which is also well understood in string theory \cite{Dymarsky:2006ve}.
Supersymmetric Wilson loops with non-trivial vacuum expectation values
are also of prime interest. The 1/2 BPS circle was understood early-on
to be described by a zero-dimensional theory - the celebrated
Hermitian matrix model of Erickson, Semenoff, and Zarembo
\cite{Erickson:2000af}. This matrix model appears to encode the object
entirely \cite{Drukker:2000rr}, including the string-side
manifestation of large representations
\cite{Drukker:2005kx,Gomis:2006sb,Gomis:2006im,Yamaguchi:2007ps,
  Hartnoll:2006ib,Hartnoll:2006is,Drukker:2006zk,Okuyama:2006jc} and
two-point functions with local operators
\cite{Berenstein:1998ij,Semenoff:2001xp,Giombi:2006de,Semenoff:2006am}.
Indeed, a recent paper \cite{Pestun:2007rz} has claimed a proof of
this result. Recently, a much larger class of supersymmetric loops
with non-trivial expectation values were discovered
\cite{Drukker:2007dw}. These loops lie on an $S^3$ and are
generically 1/16 BPS. An important subclass of those loops lie on a
great $S^2$ inside the $S^3$. It has been suggested by their
discoverers that these Wilson loops might be captured exactly by a
reduced two-dimensional model which one could describe roughly as a
perturbative pure Yang-Mills theory on $S^2$, where the
Wu-Mandelstam-Leibbrandt
\cite{Wu:1977hi,Mandelstam:1982cb,Leibbrandt:1983pj} prescription
for the regularization of the propagator is used
\cite{Drukker:2007yx,Drukker:2007qr}. We will refer to this simply
as the ``reduced 2-d model''.

The Wilson loop on $S^2$ proposed by \cite{Drukker:2007dw} is given by

\be\label{theloop}
W = \frac{1}{N} \Tr {\cal P} \exp \oint d\t \,
\left( i \,\dot x^i A_i + \e_{ijk}\, x^j \dot x^k \,M^i_I \,\Phi_I \right)
\ee

\ni where $x^i(\t)$ (where $i=1,\ldots,3$, $I=1,\ldots,6$) is a closed
path on $S^2$, and $M^i_I$ is a $3\times 6$ matrix satisfying $M
M^T=1$ and which we will take to be $M^i_i=1$ (no summation implied)
and all other entries zero. The existing evidence that this object
might be captured by a reduced 2-d model has been presented in
\cite{Drukker:2007yx} and \cite{Drukker:2007qr}. Here we will give a
short review of those results. One of the most compelling observations
is that the combined scalar and gauge field (Feynman gauge) propagator
joining two points $x$ and $y$ on the loop (the so-called
``loop-to-loop propagator'') is given by

\be\label{propagator} D_{4d} \propto \frac{g^2}{R^2} \left(
\frac{1}{2} \d_{ij} - \frac{(x-y)_i (x-y)_j}{(x-y)^2} \right),\qquad
i,j=1,2,3 \ee

\ni where $R$ is the radius of the $S^2$. This indeed is the
propagator of pure 2-d Yang-Mills in a certain gauge, with coupling
$g^2_{\text{2d}} = -g^2/(4\pi R^2)$. Using this one can prove via
Stokes theorem that for a general closed contour on $S^2$

\be\label{pert} \la W \ra = 1 + \l \,\frac{{\cal A}_1{\cal
A}_2}{{2\cal A}^2} + {\cal O}(\l^2), \ee

\ni where $\l = g^2 N$, and ${\cal A}_1$ and ${\cal A}_2$ are the two
areas of the $S^2$ bounded by the Wilson loop, while ${\cal A}$ is
their sum, the total sphere area. This result can then be compared to
that for a Wilson loop of arbitrary path in 2-d Yang-Mills on $S^2$ in
the zero-instanton sector, as calculated by Bassetto and Griguolo
\cite{Bassetto:1998sr} using the expansion of Witten
\cite{Witten:1991we,Witten:1992xu}\footnote{In the work
  \cite{Gross:1994mr}, it was shown that in summing this expansion,
  instantons are crucial for the recovery of strong coupling physics
  \cite{Douglas:1993iia}.}. Under the proposed relation between the 2-d
and 4-d coupling, that result reads\footnote{$L_n^m$ is the Laguerre
  polynomial $L_n^m(x)=1/n!\exp[x]x^{-m}(d/dx)^n (\exp[-x]x^{n+m})$.}

\be \label{prop}
\la W \ra = \frac{1}{N} L_{N-1}^1\left(-g^2 \frac{{\cal A}_1{\cal
    A}_2}{{\cal A}^2}\right)\exp \left(\frac{g^2}{2} \frac{{\cal A}_1{\cal
    A}_2}{{\cal A}^2}\right),
\ee

\ni and agrees with (\ref{pert}) to first order in $\l$. In fact the
1/2 BPS circular Wilson loop of ${\cal N}=4$ supersymmetric
Yang-Mills theory, and further, Drukker's 1/4 BPS generalization of
it \cite{Drukker:2006ga} are special cases of (\ref{theloop}). As
mentioned above, there exists a wealth of evidence (both at weak and
at strong coupling, and especially for the 1/2 BPS circle) that
these loops are described exactly by a Hermitian matrix model, whose
result for $\la W \ra$ agrees precisely with (\ref{prop}). Finally,
the authors in \cite{Drukker:2007qr} present a strong coupling
calculation of $\la W\ra$ for a Wilson loop composed of two
longitudes separated by an arbitrary angle using the AdS/CFT
correspondence. That result is also in agreement with (\ref{prop}).

In the decompactification limit $R\rightarrow \infty$, (\ref{prop})
agrees with the perturbative calculation of Staudacher and Krauth
\cite{Staudacher:1997kn}, performed by summing-up ladder diagrams in
the light-cone Wu-Mandelstam-Leibbrandt prescription for 2-d
Yang-Mills in the plane. The ``2-d reduced model'' proposed in
\cite{Drukker:2007qr} is essentially the same idea; albeit on $S^2$
rather than the plane and in a different gauge. They first give an
action on an $S^2$ parametrized by complex coordinates $z$, $\bar z$

\be
 x^i  = \frac{1}{1+z\bar z} ( z+ \bar z, -i(z- \bar z), 1-z\bar
z).
 \ee

\ni Beginning with generalized Feynman gauge with gauge parameter
$\xi = -1$ they propose the following Langragian density

\be {\cal  L} = \frac{\sqrt{g}}{g^2_{\text{2d}}} \left[\frac{1}{4}
(F^a_{ij})^2 -
 \frac{1}{2}(\nabla^i A_i^a)^2 \right]
\ee

\ni where $g$ is the determinant of the $S^2$ metric (i.e. $ds^2 = 4
dz d\bar z/(1+z\bar z)^2$). This leads to propagators for the $A_z$
and $A_{\bar z}$ fields as follows

\bsp\label{azz}
 \la A_z(z)\,A_z(w) \ra = \frac{g^2_{\text{2d}}}{\pi}
\frac{1}{(1+z\bar z)}\frac{1}{(1+w\bar w)} \frac{\bar z-\bar
w}{z-w}\\
\la A_{\bar z}(z)\,A_{\bar z}(w) \ra = \frac{g^2_{\text{2d}}}{\pi}
\frac{1}{(1+z\bar z)}\frac{1}{(1+w\bar w)} \frac{z-w}{\bar z-\bar w}
\end{split}
 \ee

 \ni which agree with (\ref{propagator}) when written in the original
 cartesian coordinates. In the decompactification limit these
 propagators coincide with the Wu-Mandelstam-Leibbrandt, light-cone
 propagators used by Staudacher and Krauth, up to a factor of 2.
 However one can change here to a light-cone gauge, setting $A_{\bar z}=0$;
 this gauge choice just results in twice the first propagator in
 (\ref{azz}). This light-cone gauge propagator takes on the form

\be D_{4d} + i D_0 \ee

\ni where $D_{4d}$ is the loop-to-loop propagator from ${\cal N}=4$
SYM in Feynman gauge (i.e. (\ref{propagator})) while $i D_0$ is a new
imaginary piece generated by the gauge transformation. Employing this
gauge affords a great simplification in Feynman diagrams since
interactions are clearly removed; one needs only consider the sum of
ladder diagrams. These might reproduce (\ref{prop}) for single Wilson
loops on $S^2$. For the connected correlator of two Wilson loops, one
can simply compare ${\cal N}=4$ SYM results to ladder diagrams.

\begin{figure}[ht]
\begin{center}
\includegraphics*[height=1.0in]{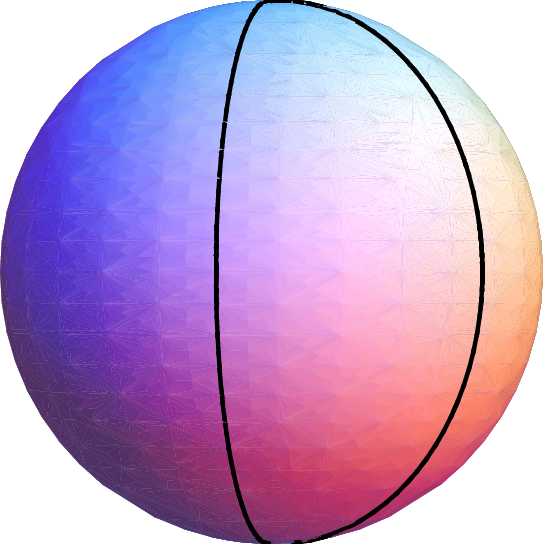}\hspace{0.3in}
\includegraphics*[height=1.0in]{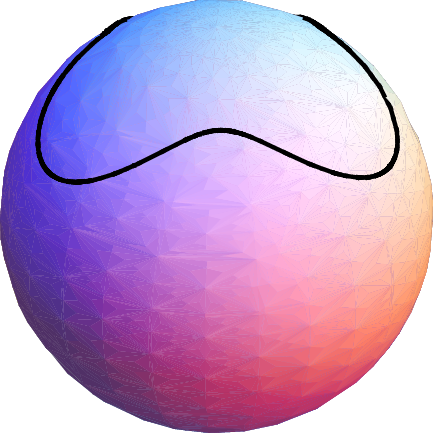}\hspace{0.3in}
\includegraphics*[height=1.0in]{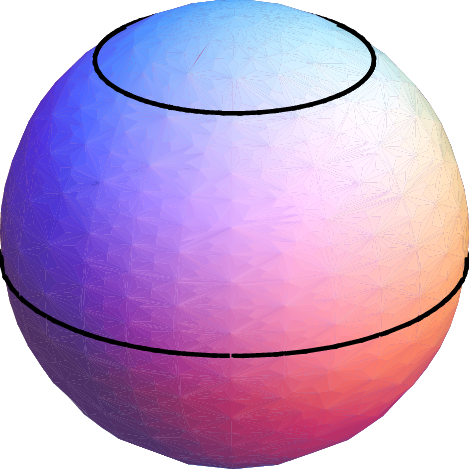}
\end{center}
\caption{The three geometries of Wilson loops on $S^2$ we consider:
  two longitudes, wavy-latitude, and two latitudes.}
\label{fig:loops}
\end{figure}

It is the purpose of this paper to explore the connection of the
Wilson loops (\ref{theloop}) to the proposed reduced 2-d model
further. We consider the vacuum expectation value (VEV) of the Wilson
loop constructed in \cite{Drukker:2007qr} consisting of two longitudes
to second order in the 't Hooft coupling. The resulting integrals
involve Feynman parameters as well as integrations over the longitudes
themselves. We find that for angles away from zero separating the
longitudes, numerical integration produces accurate results. These are
in excellent agreement with (\ref{prop}). We continue to the same
calculation for a ``wavy-latitude'': a latitude with a sinusoidal wave
of low period in the polar angle describing it, see figure
\ref{fig:loops}. Using the same techniques, we similarly find
excellent agreement with (\ref{prop}), and for a continuous range of
wave amplitudes.
We also consider the connected correlator of two distinct latitudes to
third order in the 't Hooft coupling. In this case we cannot compare
to the Bassetto and Griguolo result, as that result is valid for the
VEV of a single Wilson loop and not a connected correlator of two.
Instead we compare to the reduced 2-d model of \cite{Drukker:2007qr}
presented above, in light-cone gauge. The reduced model produces
results which are consistent with the result from planar ${\cal N} =
4$ SYM at leading order (second order) in the 't Hooft coupling.
At the next order, i.e. third order in the 't Hooft coupling,
we attempt to make a comparison in the limit where the latitudes
becomes coincident. We find that we are unable to calculate the result
at the necessary order in the separation of the latitudes to test for
a match with the reduced model, but the result also does not preclude
the existence of such a match.

The structure of the paper is as follows. In section \ref{<w>} we
calculate the VEV of single Wilson loops; we consider the case of two
longitudes and of a wavy latitude. In section \ref{sec:cc} we compare
the connected correlator of two latitudes, as calculated in ${\cal
  N}=4$ SYM to the expectation from the 2-d reduced model in
light-cone gauge. We conclude with a discussion of the results in
section \ref{sec:end}. The details of the calculations, which are very
complicated, have been included in the appendices. As this manuscript
was being readied for publication \cite{Bassetto:2008yf} appeared
which has some overlap with section \ref{sec:long}.

\subsection*{Notes concerning v4}

In the earlier versions of this manuscript an error was present in the
analysis of the coicident limit of the correlator calculation. The
original claim was that there was a mismatch between the
two-dimensional Yang-Mills result and the one coming from ${\cal N}=4$
SYM in the case of the correlator. We have discovered that the error
lay in taking the coincident limit - having fixed the error we find
that we cannot get results at the same order in the separation between
the latitudes which were advertised in previous versions.

\section{Calculations of $\la W \ra$ at ${\cal O}(\l^2)$}
\label{<w>}

We consider the VEV of a Wilson loop of the variety (\ref{theloop}).
As explained in the introduction, at ${\cal O}(\l)$ these loops have
been proven to be captured by (\ref{prop}). We would like to
understand whether this agreement persists at the next order in
perturbation theory. A two-loop calculation was performed for the 1/2
BPS circle in \cite{Erickson:2000af}; we follow that calculation
closely and refer the reader there for conventions and notation. We
use the Euclidean action of ${\cal N}=4$ SYM in Feynman gauge and
dimensional regularization.

\begin{figure}[ht]
\begin{center}
\includegraphics*[bb= 11 9 275 120,height=1.0in]{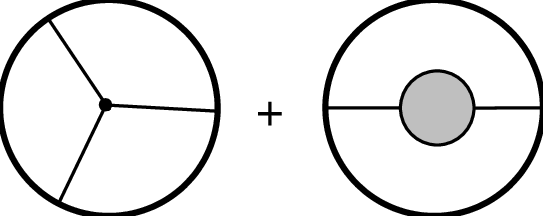}
\end{center}
\caption{The two-loop, non-ladder/rainbow diagrams contributing to
  $\la W \ra$. The Wilson loop is indicated by the outer circle.
  Internal solid lines refer to scalar and gauge fields, while the
  greyed-in bubble represents the one-loop correction to the
  propagator.}
\label{fig:iv}
\end{figure}

There are three types of diagrams contributing to $\la W \ra$ at
${\cal O}(\l^2)$. The simplest are the rainbow/ladder graphs - those
graphs without interaction vertices. The next contributions come
from diagrams with interaction vertices, these are shown
schematically in figure \ref{fig:iv}.
In what follows we will assume smooth Wilson loop contours; the case
of the two longitudes will be slightly different. We may generalize
eq. (13) of \cite{Erickson:2000af}, which gives the contribution from
the diagram on the left in figure \ref{fig:iv}. In keeping with their
notation, we call this quantity $\S_3$

\be \S_3 = -\frac{\l^2}{4} \oint d\t_1 \,d\t_2 \,d\t_3 \,
\e(\t_1\,\t_2\,\t_3)\, D(\t_1,\t_3)\, \dot x_2 \cdot \p_{x_1}
G(x_1,x_2,x_3) \ee

\ni where\footnote{The symbol $\e(\t_1\,\t_2\,\t_3)$ refers to
  antisymmetric path-ordering. It is given by +1 for $\t_1 > \t_2 >
  \t_3$ and is totally antisymmetric in the $\t_i$.} we have used
$D(\t_1,\t_3)$ to refer to the numerator of the loop-to-loop
propagator, i.e. in our case $D(\t_1,\t_2) = (\dot x_1 \cdot \dot x_2)
( x_1 \cdot x_2 - 1) - (x_1 \cdot \dot x_2)(x_2 \cdot \dot x_1) $,
while the function $G$ is as defined in \cite{Erickson:2000af}

\bsp G(x_1,x_2,x_3) = \frac{\G(2\o-3)}{2^6 \pi^{2\o}} \int_0^1
d\alpha\,d\beta\,d\gamma\,(\alpha\beta\gamma)^{\omega-2}
\delta(1-\alpha-\beta-\gamma)\,\\
\times \frac{1} {\bigl[ \a\b(x_1-x_2)^2 + \b\g(x_2-x_3)^2 +
\a\g(x_1-x_3)^2 \bigr]^{2\o-3}}
\end{split}
\ee

\ni where the number of dimensions is given by $d=2\o$, so that the
physical dimension is at $\o = 2$. Using the fact that\footnote{This
is the relation which must be modified for curves which are
piecewise defined.}

\be \oint d\t_1 \,d\t_2 \,d\t_3 \,\frac{d}{d\t_1} \Biggl(
\e(\t_1\,\t_2\,\t_3)\, D(\t_1,\t_3)\,G(x_1,x_2,x_3)  \Biggr) = 0 \ee

\ni one may prove that

\be\label{bbb} \frac{\l^2}{2} \oint d\t_1 d\t_3 \,
\frac{D(\t_2,\t_3)}{G|_{\t_1 =
    \t_2}} =
-\frac{\l^2}{4} \oint d\t_1\,d\t_2\, d\t_3\, \e(\t_1\,\t_2\,\t_2)\,
\p_{\t_1} \biggl( D(\t_1,\t_3)\, G \biggr). \ee

\ni In fact, as shown in \cite{Erickson:2000af}, on the physical
dimension, the LHS of the expression (\ref{bbb}) (which is
divergent) reduces to exactly minus the contribution of the diagram
pictured on the right of figure \ref{fig:iv}. The sum of the two
diagrams is therefore given by (calling the contribution of the
second diagram $\S_2$)

\be\label{2l} \S_3 + \S_2 = -\frac{\l^2}{4} \oint d\t_1 \,d\t_2
\,d\t_3 \, \e(\t_1\,\t_2\,\t_3)\, \Biggl[ D(\t_1,\t_3)\,\dot x_2
\cdot \p_{x_1} G - \p_{\t_1} \biggl(D(\t_1,\t_3)\,G \biggr) \Biggr]
\ee

\ni which for the 1/2 BPS circle \cite{Erickson:2000af}, and for the
latitude \cite{Drukker:2006ga} is easily proven to be zero. As
long as the Wilson loop under consideration is finite at one-loop, i.e.

\be
\oint d\t_1\, d\t_2\, \frac{D(\t_1,\t_2)}{(x_1-x_2)^2} = \text{finite}
\ee

\ni it also easy to see that (\ref{2l}) is finite. We will discuss this
point further in section \ref{sec:disc<w>}.

Our strategy is to evaluate the rainbow/ladders and the quantity
(\ref{2l}) using numerical integration, and to compare to the
expectation from (\ref{prop}). Expanding that expression in the
large-$N$, small-$\l$ limit, one finds

\bsp\label{2dYM} \la W \ra
= 1 + \frac{\hat \l}{8} {\cal A}_1(4\pi -{\cal A}_1) + \frac{\hat
\l^2}{192} \left( {\cal A}_1(4\pi -{\cal A}_1)\right)^2+\ldots
\end{split}
\ee

\ni where we have defined $\hat\l \equiv \l/(4\pi^2)$ and where ${\cal
  A}_1$ is either of the areas enclosed by the Wilson loop on $S^2$.

\subsection{Two longitudes}
\label{sec:long}

\begin{figure}[ht]
\begin{center}
\includegraphics*[bb= 33 28 195 185,height=1.5in]{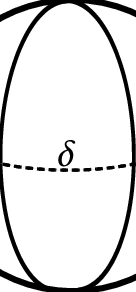}
\end{center}
\caption{A Wilson loop composed of two longitudes.} \label{fig:twolongs}
\end{figure}

We consider the Wilson loop defined by (\ref{theloop}) consisting of
two longitudes separated by an azimuthal angle $\d$ on $S^2$, as
pictured in figure \ref{fig:twolongs}. 
This loop was first constructed in
\cite{Drukker:2007qr} and it is relatively straightforward to prove
that it is indeed captured by (\ref{prop}) to first order in the 't
Hooft coupling directly. The longitudes are given by

\bsp
&x^i = (\sin t,\, 0,\, \cos t),\quad 0 \leq t < \pi\\
&x^i = (-\cos \d \sin t,\, -\sin \d \sin t,\, \cos t),\quad \pi \leq
t < 2\pi
\end{split}
 \ee

\ni where the first longitude couples to the scalar field $\Phi_2$,
and the second to $-\Phi_2 \cos \d + \Phi_1 \sin \d$. The combined
gauge field and scalar propagator joining two points on the same
longitude is a constant $\l/(4\pi^2)\times 1/2 = \hat \l /2$, while
that joining the two longitudes is given by

\be P(t_1,t_2) = \hat \l\,\frac{-\dot x_1 \cdot \dot x_2 - \cos \d}
{2(1-x_1 \cdot x_2)} = \hat\l\,\frac{\cos\d \cos t_1 \cos t_2 - \sin
t_1 \sin t_2 - \cos\d}{2(1 + \cos \d
  \sin t_1 \sin t_2 -\cos t_1 \cos t_2 ) }.
\ee

\begin{figure}
\begin{center}
\includegraphics*[bb= 0 244 237 445,height=1.in]{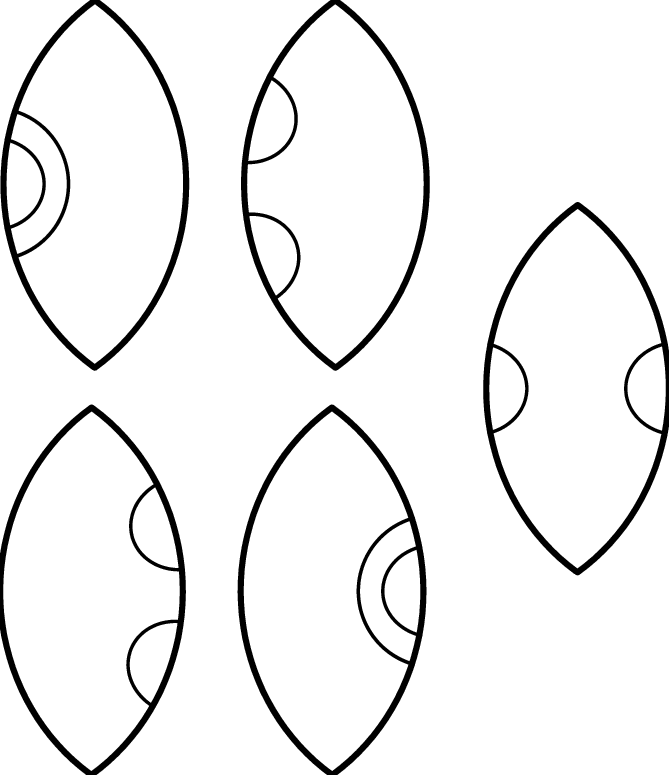}
\includegraphics*[bb= 10 44 242 240,height=0.95in]{twoloop.eps}
\includegraphics*[bb= 250 140 375 350,height=1.0in]{twoloop.eps}
\end{center}
\caption{A subset of the two-loop diagrams.} \label{fig:twoloop}
\end{figure}

\ni We begin with those rainbow/ladder graphs which do not involve
the propagator $P(t_1,t_2)$; these are pictured in figure
\ref{fig:twoloop}.
We find that these diagrams yield the following

\bsp\label{noP} \frac{\hat \l^2}{4} \Biggl[ &2 \int_{0}^{\pi} dt_1
\int_0^{t_1} dt_2
  \int_0^{t_2} dt_3 \int_0^{t_3} dt_4\,\left(\frac{1}{2}\right)^2
+  2 \int_{\pi}^{2\pi} dt_1 \int_{\pi}^{t_1} dt_2
  \int_{\pi}^{t_2} dt_3 \int_{\pi}^{t_3}
  dt_4\,\left(\frac{1}{2}\right)^2\\
&+ \int_{\pi}^{2\pi} dt_1 \int_{\pi}^{t_1} dt_2
  \int_{0}^{\pi} dt_3 \int_{\pi}^{t_3}
  dt_4\,\left(\frac{1}{2}\right)^2 \Biggr] =
\frac{\hat \l^2}{16} \left[ (2+2)\cdot\frac{\pi^4}{4!} +
  \left(\frac{\pi^2}{2!}\right)^2 \right] =
\frac{5\hat \l^2}{192} \pi^4
\end{split}
\ee

\begin{figure}[ht]
\begin{center}
\includegraphics*[bb= 30 38 495 230,height=0.9in]{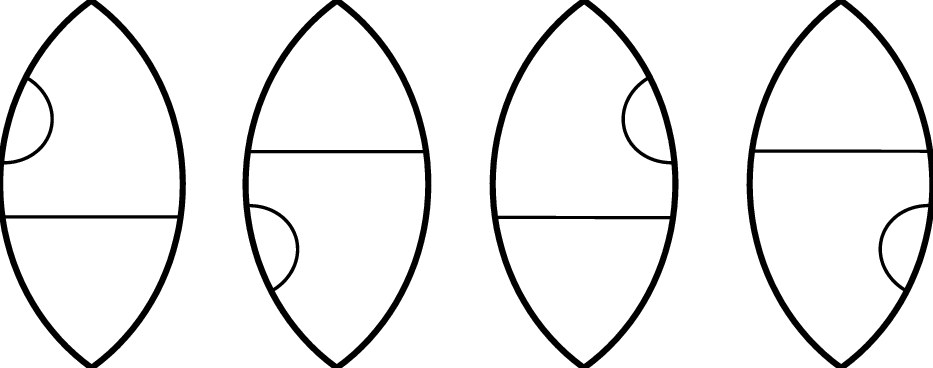}
\includegraphics*[bb=33 41 136 231,height=0.9in]{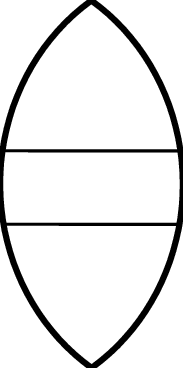}
\end{center}
\caption{A (different) subset of the two-loop diagrams.}
\label{fig:twoloopb}
\end{figure}

\ni where the leading factor of $1/4$ comes from the traces over
gauge group matrices, while the $1/4!$ which comes from the
expansion of the Wilson loop to fourth order has been eliminated by
the $4!$ equivalent orderings of the fields in that expansion. The
next class of two-loop rainbow/ladder diagrams contain the
$P(t_1,t_2)$ propagator and are pictured in figure
\ref{fig:twoloopb}.

\ni We find the result for these diagrams to be

\bsp\label{oneP} \L_2 \equiv \frac{\hat \l^2}{2}  \int_{\pi}^{2\pi}
dt_1 \int_0^{\pi} dt_2&
  \int_0^{t_2} dt_3 \int_0^{t_3} dt_4\,\left(\frac{1}{2}\right)\biggl(
P(t_1,t_4) + P(t_1,t_2) \biggr)\\
&+\frac{\hat\l^2}{4}   \int_{\pi}^{2\pi} dt_1 \int_{\pi}^{t_1} dt_2
  \int_0^{\pi} dt_3 \int_0^{t_3} dt_4\,
P(t_1,t_4)\, P(t_2,t_3).
\end{split}
\ee

\ni There are two checks which we can make on the sum of two-loop
rainbow/ladders. The first is at $\d = 0$ where the longitudes lie
atop one another with opposite orientation. Here the result should be
zero, and is. The second is at $\d=\pi$ where the longitudes
degenerate to a great circle. Here the result should match that of the
1/2 BPS circle, since there internal vertex diagrams cancel
\cite{Erickson:2000af}. One can check that this test is also passed.

\begin{figure}[h]
\begin{center}
\includegraphics*[bb= 60 265 490 685,height=2.9in]{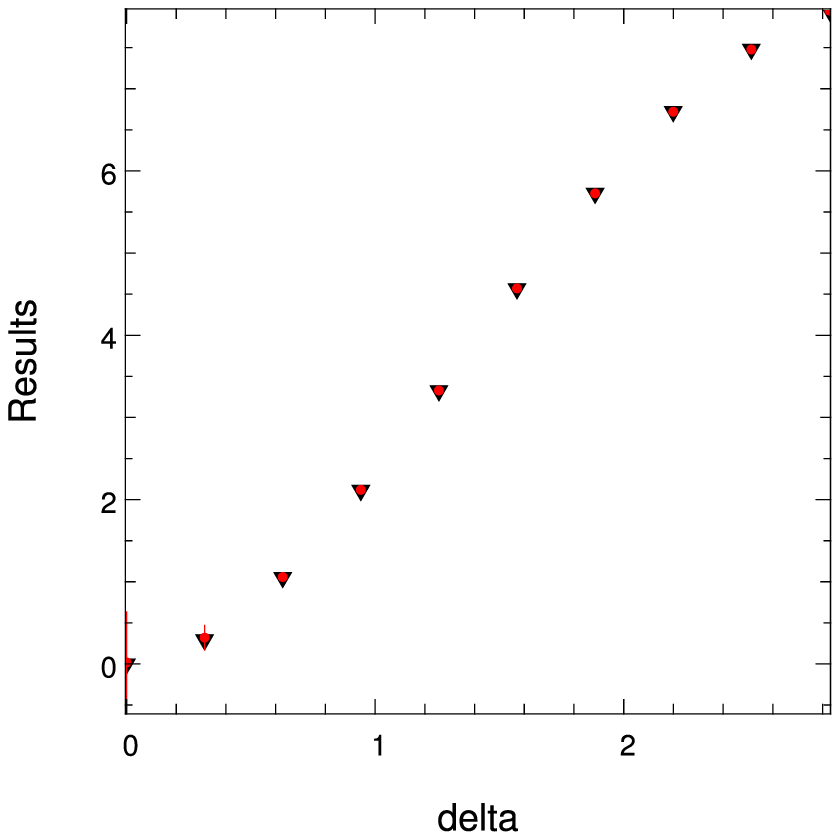}
\includegraphics*[bb= 60 265 490 685,height=2.9in]{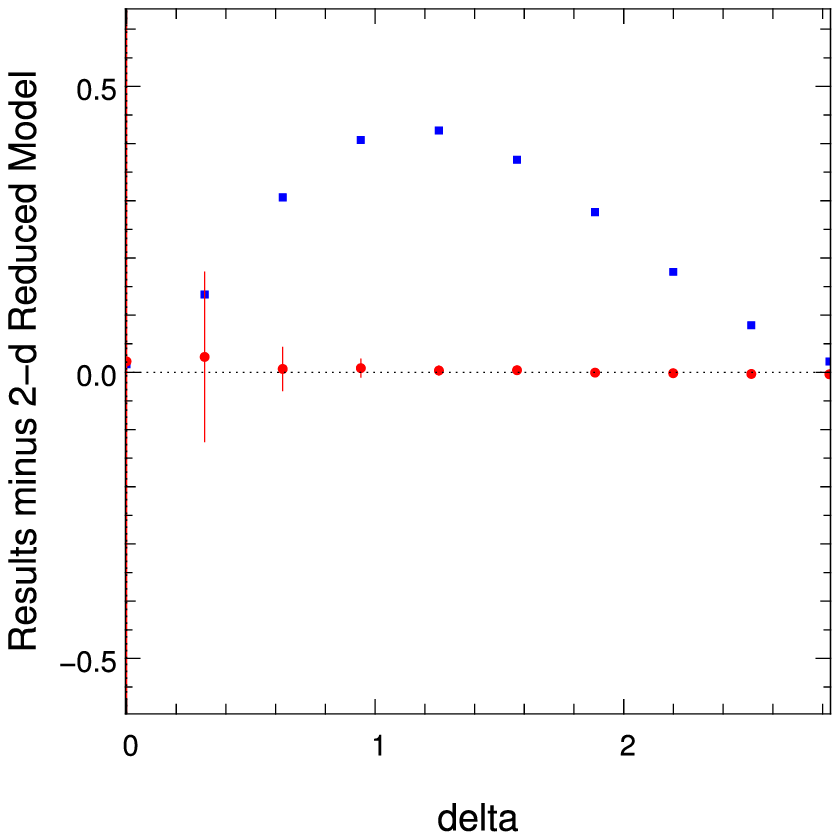}
\caption{
  Two-loop results for a Wilson loop composed of two longitudes
  ($\hat \l$ is set to 1). In red dots the result of numerical
  integration is shown. In black triangles the expectation from
  (\ref{prop}) is plotted. On the right data including the result
  from only rainbow/ladder diagrams (blue squares) are plotted with the
  expectation from (\ref{prop}) subtracted.}
\label{fig:blahah}
\end{center}
\end{figure}

The expectation from (\ref{2dYM}) at two-loop order is easily seen to
be $\hat \l^2\, \d^2(2\pi-\d)^2/12$. It is interesting to ask whether
or not the sum of two-loop rainbow/ladder diagrams is already
proportional to $\d^2(2\pi-\d)^2$, even without the contribution of
the internal vertex diagrams. Due especially to the last integral in
(\ref{oneP}), we need to resort to numerical integration in order to
answer this question. As we will see the answer is no. The internal
vertex diagrams, however, give a finite contribution which together
with the rainbow/ladders, reproduces the prediction from (\ref{prop}).
Due to the fact that this Wilson loop is piecewise defined, the
interacting diagrams and their divergence cancellation is more subtle
than that presented at the start of this section. We have relegated
the details to appendix \ref{sec:longitudes}. We find the following
result for the finite remainder after the divergence cancellation

\bsp\label{L3} \L_3 = -\frac{\hat \l^2}{16} \int_0^1 d\a\, d\b\,
d\g\, \d(1-\a-\b-\g) \Biggl[  \int_\pi^{2\pi} d\t_1 \int_0^\pi d\t_2
\int_0^{\pi}
d\t_3 \, \e(\t_2 \,\t_3) \, \frac{B_1+B_2+B_3}{\D^2}\\
- \int_\pi^{2\pi} d\t_1 \int_0^\pi d\t_2
\frac{(1+\s)(2+c_1+c_2)}{[\a\b(1+\s s_1 s_2 - c_1 c_2) + \b\g(1+c_2)
+ \a\g(1+c_1)]} \Biggr]
\end{split}
\ee

\bsp B_1+B_2+B_3 = &\a\g (\s^2-1) \left[ 2 s_1 (c_3 - c_2) - s_1 c_1
(1-\cos \t_{23}) \right]\\
&+ \a\g(\s+1)(s_2-s_3)(c_3-c_1)\\
&+\a\g(\s+1)\left[\sin\t_{13}^+ - \sin \t_{12}^+ +\sin\t_{23} \right]\\
&+\a\g(\s+1) \sin\t_{23}(1-\cos\t_{13}^+) +\b\g (\s+1) c_1 s_3
(1-\cos\t_{23})
\end{split}
\ee

\ni where we have introduced some shorthand $\s \equiv \cos\d$, $c_i
\equiv \cos \t_i$, $s_i \equiv \sin \t_i$, $\t_{ij}\equiv \t_i -
\t_j$, $\t_{ij}^+ \equiv \t_i + \t_j$, and

\bsp \D = &\a\b ( 1+ \s \sin \t_1 \sin \t_2 - \cos \t_1 \cos \t_2 )
+
\b \g (1 - \cos \t_{23})\\
& + \a \g ( 1 + \s \sin \t_1 \sin \t_3 - \cos \t_1 \cos \t_3 ).
\end{split}
\ee

We have evaluated the complete result $5\hat \l^2 \pi^4/192 + \L_2 +
\L_3$ via numerical integration. The results are shown in figure
\ref{fig:blahah} for a range of opening angles $\d$ as red dots with
estimated error bars.  Also plotted as black triangles is the
expectation from (\ref{prop}), i.e. $\hat \l^2\, \d^2(2\pi-\d)^2/12$.
On the right the results, including the rainbow/ladder contribution
alone (i.e.  $5\hat \l^2 \pi^4/192 + \L_2$) are plotted with the
expectation from (\ref{prop}) subtracted. It is clear both that the
rainbow/ladders fail to reproduce the expectation from (\ref{prop}),
and that the addition of $\L_3$, at least for angles $\d$ away from
$\d=0$, reproduces them excellently. As $\d=0$ is approached the
numerical integration is no longer reliable (as evidenced by the
growing error bars). The reasons for this are discussed in section
\ref{sec:disc<w>}. We also note from (\ref{L3}) that $\L_3$ vanishes
exactly for $\d = \pi$ when the longitudes degenerate to a circle;
this is a consistency check against the known vanishing of interacting
diagrams for the 1/2 BPS circle \cite{Erickson:2000af}.

\subsection{Wavy latitudes}

We now consider (\ref{theloop}) using a class of closed contours we
refer to as ``wavy latitudes''. They are given by

\be\label{cont} \vec x(\t) = \bigl( \sin \theta(\t) \cos \t,\, \sin
\theta(\t) \sin\t,\,\cos \theta(\t) \bigr), \qquad \theta(\t) =
\theta_0 + A\cos n\t \ee

\begin{figure}[ht]
\begin{center}
\includegraphics*[bb= 96 305 486 671,height=2.in]{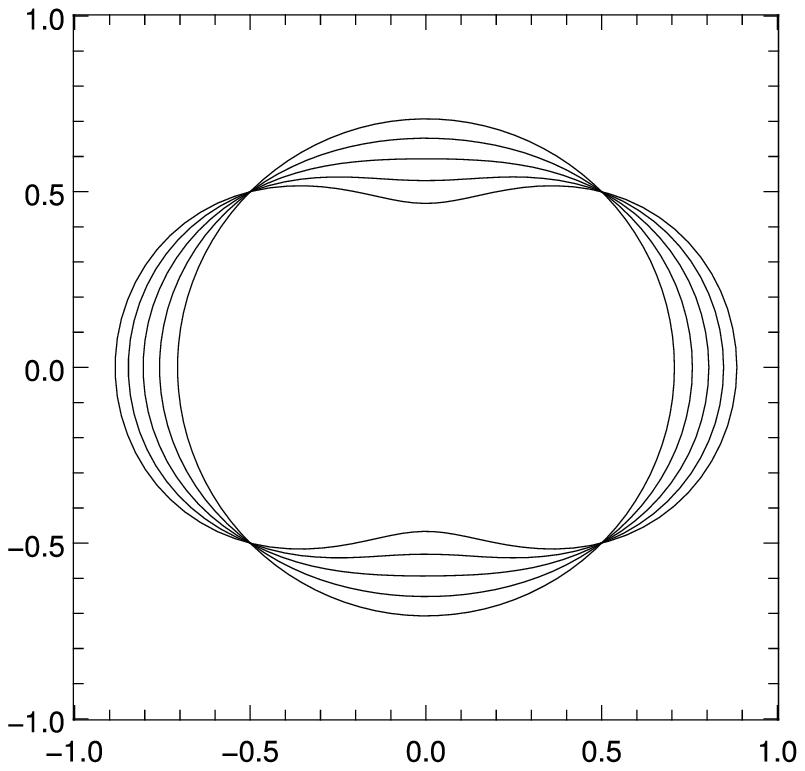}
\includegraphics*[bb= 96 305 486 671,height=2.in]{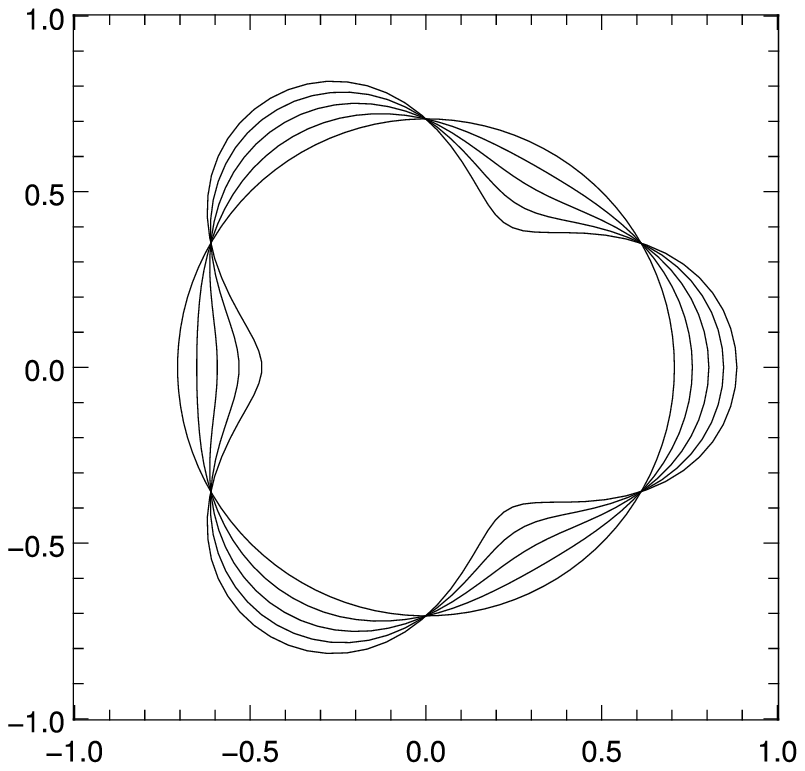}
\end{center}
\caption{The contours (\ref{cont}) are plotted from the view-point
  straight down the north pole of the sphere (flat projection). Here $\theta_0=\pi/4$
  while $A$ ranges from 0 to 0.3. On the left $n$ has been set to 2,
  on the right $n=3$.}
\label{fig:c1}
\end{figure}

\ni where $n$ is an integer. For $A=0$ these loops reduce to the
latitudes which were shown in \cite{Drukker:2007dw} to be
essentially the same (via a conformal transformation) as the 1/4 BPS
circle of Drukker \cite{Drukker:2006ga}, and for which the 1/2 BPS
circle is a special case. In figure \ref{fig:c1}, we have plotted
the curves for $\theta_0 = \pi/4$, and $A$ ranging from 0 to 0.3 for
the cases $n=2,3$. The viewpoint is straight down the north pole of
the sphere, i.e. the contours have been (flatly) projected into the
$x_1$-$x_2$ plane.
\begin{figure}
\begin{center}
\includegraphics*[bb= 70 265 480 675,height=2.25in]{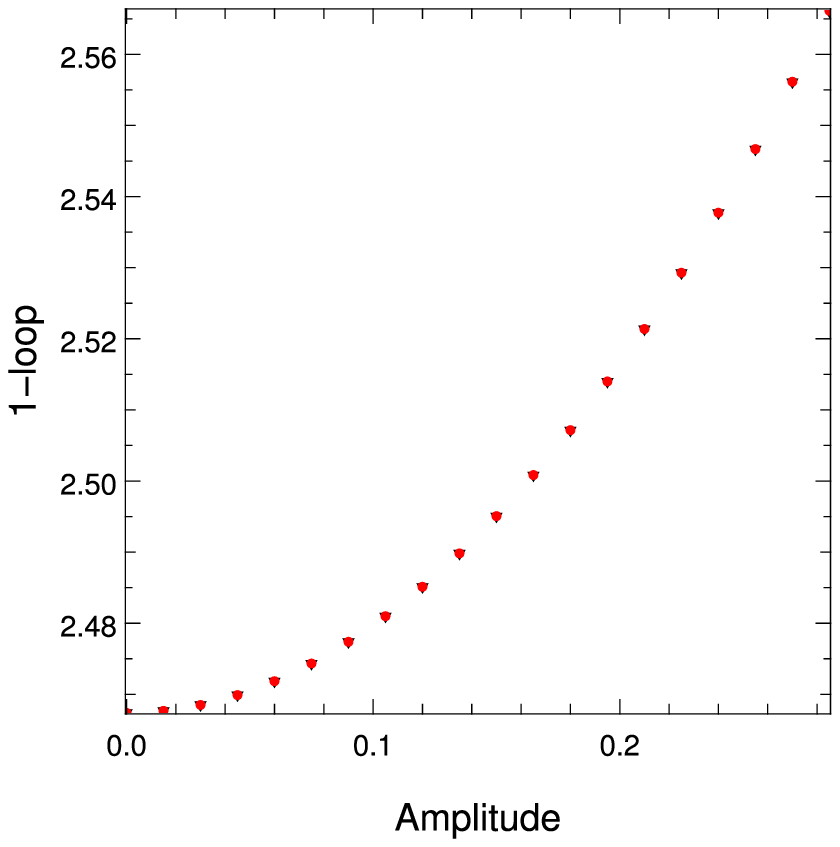}
\end{center}
\caption{The coefficient of $\hat \l$ from (\ref{2dYM})
  is plotted as black triangles for the wavy
  latitude with $\theta_0=\pi/4$ and ``amplitude'' $A$ ranging from 0
  to 0.3. Also plotted is the analogous term from ${\cal N}=4$ SYM
  perturbation theory (red dots). As they are guaranteed to by the
  results of \cite{Drukker:2007yx}, the data agree excellently.}
\label{fig:1l}
\end{figure}
\begin{figure}[!t]
\begin{center}
\includegraphics*[bb= 70 265 480 675,height=2.25in]{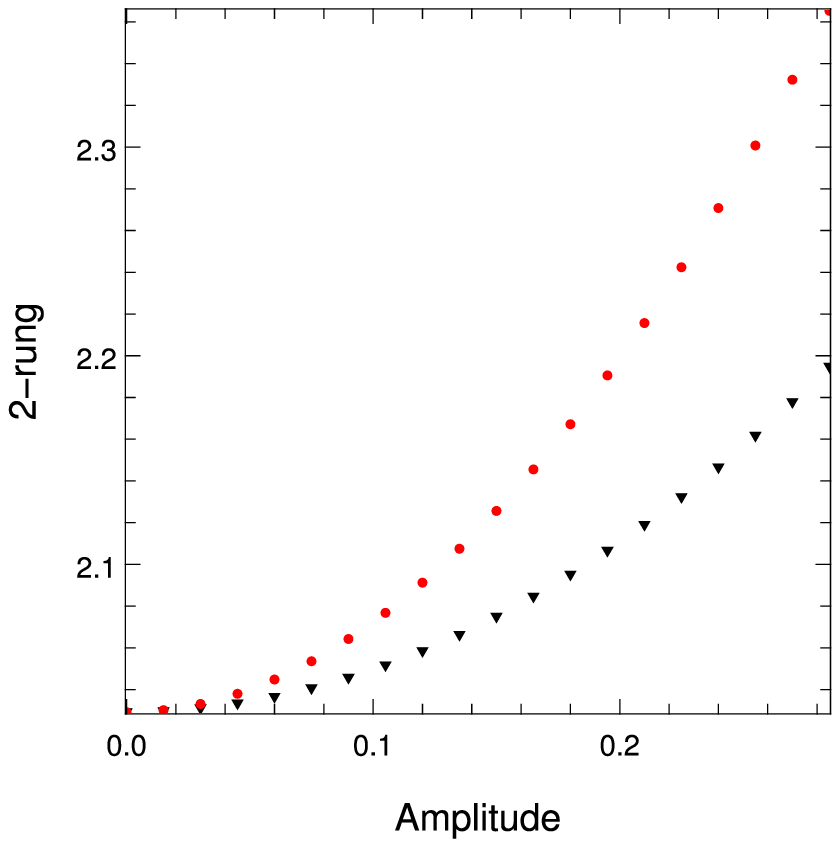}
\includegraphics*[bb= 70 265 480 675,height=2.25in]{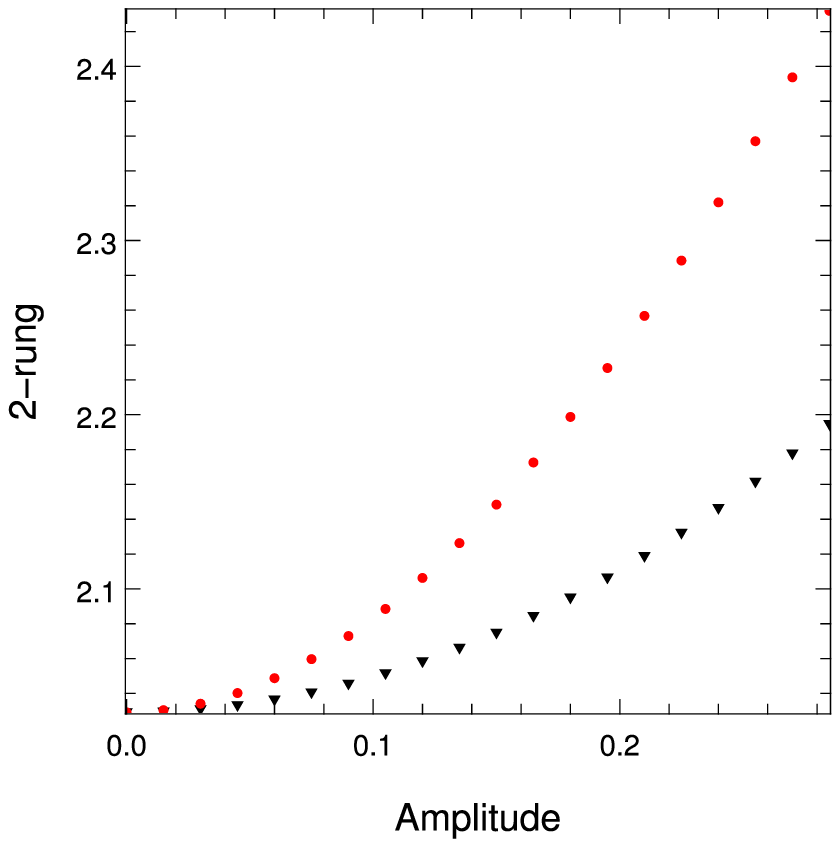}
\includegraphics*[bb= 70 265 480 675,height=2.25in]{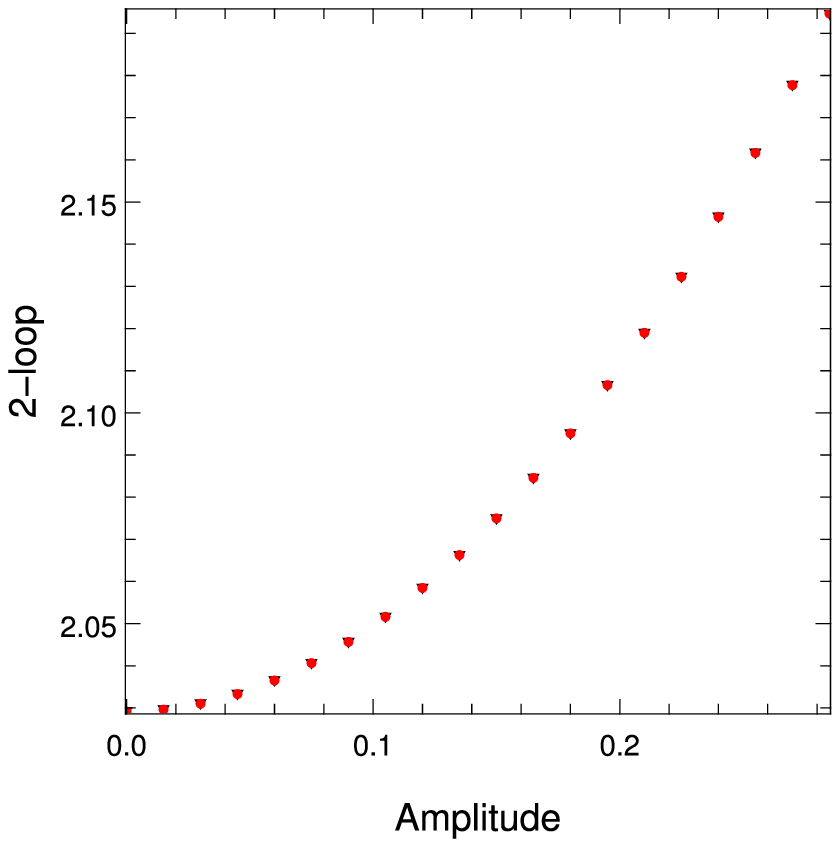}
\includegraphics*[bb= 70 265 480 675,height=2.25in]{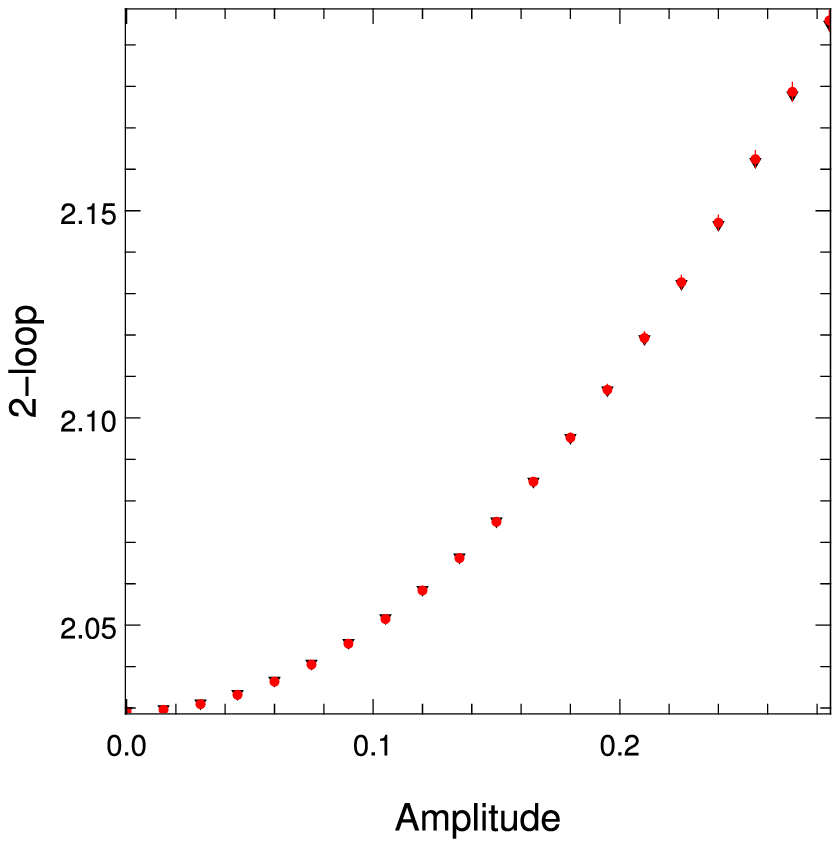}
\end{center}
\caption{In the top two graphs, the ``two-rung'' contribution
  $\S_1/\hat \l^2$ (see (\ref{tworung})) is plotted as red dots for
  the contours (\ref{cont}) with $\theta_0=\pi/4$, ``amplitude'' $A$
  ranging from 0 to 0.3, and for $n=2$ on the left and $n=3$ on the
  right. Also plotted, as black triangles, is the expectation from (\ref{2dYM}).
  In the bottom two graphs, we have
  replaced $\S_1 \rightarrow \S_1 + \S_2 + \S_3 $, i.e. the full
  two-loop result; the agreement with (\ref{2dYM}) is excellent.}
\label{fig:rungs}
\end{figure}
The rainbow/ladder contribution is given by

\be\label{tworung} \S_1 = \frac{\hat \l^2}{4} \int_0^{2\pi} d\t_1
\int_0^{\t_1} d\t_2 \int_0^{\t_2} d\t_3 \int_0^{\t_3} d\t_4 \Bigl[
Q(\t_1,\t_4) Q(\t_2,\t_3) + Q(\t_1,\t_2)Q(\t_3,\t_4) \Bigr] \ee

\ni where $Q(\t_1,\t_2)$ is defined by the integrand in
(\ref{oneloop}). We call this contribution the ``two-rung
contribution''. At ${\cal O}(\l)$, there is no need to verify
agreement of the wavy latitudes with (\ref{2dYM}), as this agreement
can already be proven for a general contour as explained in the
introduction. That being said, we may continue with the one-loop
analysis anyways, as it serves as a warm-up to the two-loop analysis
which follows. Expanding (\ref{theloop}) to leading order in the 't
Hooft coupling $\l$, we find

\be\label{oneloop} \la W \ra = 1 + \frac{\hat \l}{4} \int d\t_1 \int
d\t_2 \, \frac{(\dot x_1 \cdot \dot x_2) ( x_1 \cdot x_2 - 1) - (x_1
\cdot \dot x_2)(x_2 \cdot \dot x_1) } { 2\,(1-x_1\cdot x_2) } \ee

\ni where $x_i = \vec x(\t_i)$, we have used the fact that $x_i^2=1$,
and we have defined $\hat \l \equiv \l/(4\pi^2)$. It is not
particularly illuminating to substitute the expression for the wavy
latitude (\ref{cont}) into this expression. Instead we note that for
$A < \theta_0$ (at $A=\theta_0$ the contour self-intersects and thus
develops cusps) the expression (\ref{oneloop}) may be integrated
numerically to high accuracy. The expectation from (\ref{2dYM}),
requires the evaluation of

\be {\cal A}_1 = \int_0^{2\pi} d\t\, \bigl( 1 - \cos (\theta_0 +
A\cos n\t) \bigr). \ee

\ni This integral also requires numerical integration, however it may
be evaluated with extremely high accuracy. In figure \ref{fig:1l} we
have plotted the coefficients of $\hat \l$ from expressions
(\ref{oneloop}) and (\ref{2dYM}) for $\theta_0=\pi/4$ and the
``amplitude'' $A$ ranging from 0 to 0.3. The data lie on top of one
another, and the error bars lie within the data points\footnote{In
  these expressions there is no dependence on $n$.  }.

In figure \ref{fig:rungs} we show the numerical evaluation of the
two-rung contribution (see (\ref{tworung})) $\S_1/\hat\l^2$ for
$\theta_0=\pi/4$, $A$ ranging from 0 to 0.3, and for $n=2,3$. Also
plotted is the coefficient of $\hat \l^2$ expected from (\ref{2dYM}).
It is clear that the two-rung diagram alone does not agree with
(\ref{2dYM}), except in the trivial case $A=0$ when the regular
latitude is recovered. Also in figure \ref{fig:rungs}, in the bottom
two graphs, we show the same analysis, however this time adding the
contribution from $\S_2 + \S_3$ (see (\ref{2l})). It is seen that
within numerical accuracy, which is excellent, there is agreement with
the expectation from (\ref{2dYM}).

\subsection{Comments on numerical accuracy}
\label{sec:disc<w>}

The mechanism whereby the divergence present in (\ref{2l}) cancels
was discussed in \cite{Semenoff:2001xp}. The divergence is found by
setting the Feynman parameter $\g$ to zero. One then
finds\footnote{The divergent $\a$ integral represents an integrable
  singularity for the other Feynman parameter $\b$.} 

\be
\left( \S_2 + \S_3\right)_{\g=0} 
= -\frac{\l^2}{4} \int_0^1 \frac{d\a}{\a(1-\a)} \oint d\t_1 \, d\t_2\,d\t_3
\,\e(\t_1\,\t_2\,\t_3)\,\left(\p_{\t_2} + \p_{\t_1}\right)\left(
  \frac{D(\t_1,\t_3) }{(x_1-x_2)^2} \right)
\ee

\ni where the derivative in $\t_2$ comes from the first term in
(\ref{2l}) and the derivative in $\t_1$ from the second. Migrating
these derivatives to the path ordering symbol via integration by parts,
equal and opposite factors of $\d(\t_1 - \t_2)$ are obtained. Thus in
the $\t_1$-$\t_2$ integration there are logarithmic divergences which
cancel between the first and second term. By exploiting the symmetries
of the integration in (\ref{2l}) one can express the integrand such
that it is manifestly zero for the case of the 1/2 BPS circle. When a
small deformation such as the amplitude $A$ for the wavy latitude is turned
on, the compensating logarithmic divergences just described become
present, but are weighted by a small number which doesn't compete with
the rest of the integral. For a large enough deformation however,
the weighting is competitive and the error stemming from the numerical
integration's inability to reliably cancel-out non-converging regions
becomes significant. Although slightly modified due to its piecewise
definition, the same comments apply to the case of the two
longitudes. This is why we have been unable to obtain reliable results
when $\d$ is near zero. Analyzing the wavy latitudes for larger $n$ or
$A$ similarly leads to poor convergence.


\section{Connected correlator}
\label{sec:cc}

\begin{figure}[ht]
\begin{center}
\includegraphics*[bb= 46 46 205 205,height=1.5in]{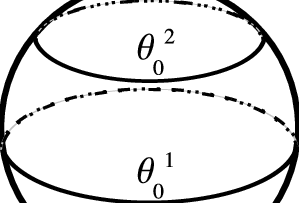}
\end{center}
\caption{Two distinct Wilson loops given by latitudes at polar angles
  $\TO^1$ and $\TO^2$.}
\label{fig:twolats}
\end{figure}

At a given order in perturbation theory, it is generally simpler to
calculate a connected correlator of two Wilson loops as compared to
the VEV of a single loop. This fact was exploited for the 1/2 BPS
circle in \cite{Plefka:2001bu,Arutyunov:2001hs} to check the matrix
model conjecture \cite{Erickson:2000af,Drukker:2000rr} to third order
in the 't Hooft coupling. We have therefore computed the connected correlator of two
Wilson loops of the variety (\ref{theloop}), given by two distinct
latitudes at polar angles $\TO^1$, $\TO^2$ on $S^2$, see figure
\ref{fig:twolats}.
The result is
compared, in the limit that the latitudes are coincident, with the
computation performed using the reduced 2-d model in light-cone gauge,
where there are only ladder diagrams. The latter result is
proportional to $h^2$, where  $h = \cos\TO^1 -
\cos\TO^2$. Unfortunately we are only able to calculate the SYM result 
at ${\cal O}(h^0)$, where we find zero. This leaves open the
possibility of further cancellations down to ${\cal O}(h^2)$, where a
match to the 2-d theory might indeed be found.

As discussed in the introduction, the reduced 2-d model light-cone
gauge propagator joining the two latitudes has the following
structure

\be
D_{2d} = D_{4d} + i D_0
\ee

\ni where $D_{4d}$ is the combined gauge and scalar field propagator
joining the latitudes in ${\cal N}=4$ supersymmetric Yang-Mills
theory in four dimensions, while $D_0$ is an extra piece (here
proportional to the difference in polar angles, i.e. $h$). Working
with gauge group $SU(N)$, and in the large-$N$ limit, it is trivial
to show equivalence between the connected correlator in the 2-d and
4-d theories at order $\l^2$. This is because in both cases, only
the 2-rung ladder diagram
\begin{center}
\includegraphics*[bb= 0 0 150 130,height=.75in]{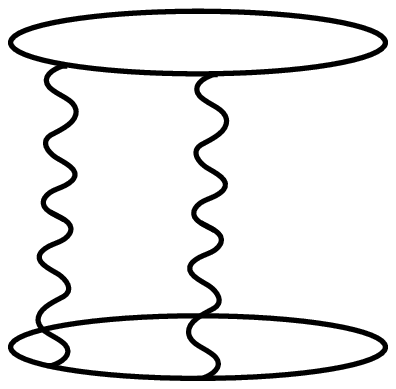}
\end{center}
contributes. Because of the form of $D_0$, it is then straight-forward
to see that its presence integrates to zero. The real test comes at
the next order in the 't Hooft coupling. At this level one can show
that, should the reduced 2-d model capture the physics
\[
\begin{minipage}[bottom]{1in}
\includegraphics*[bb= 0 0 180 130,height=.75in]{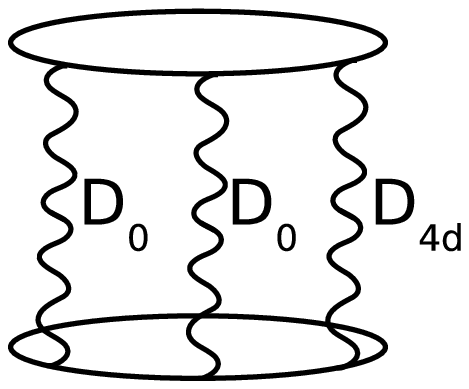}
\end{minipage}\,
= \begin{minipage}[bottom]{1in}
\includegraphics*[bb= 0 0 150 130,height=.75in]{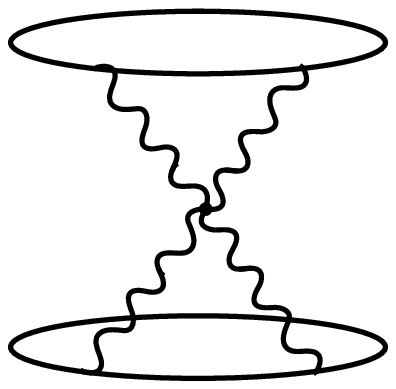}
\end{minipage}+
\begin{minipage}[bottom]{1in}
\includegraphics*[bb= 0 0 150 130,height=.75in]{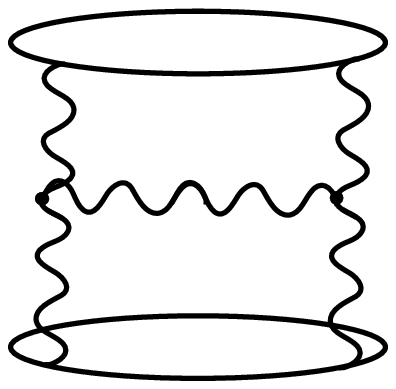}
\end{minipage}+
\begin{minipage}[bottom]{1in}
\includegraphics*[bb= 0 0 150 130,height=.75in]{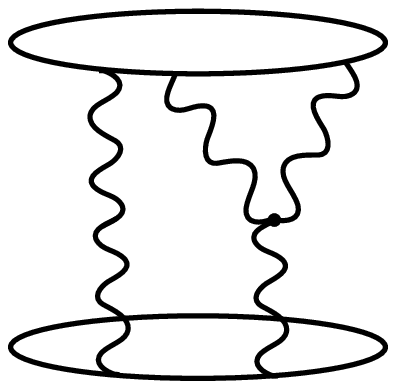}
\end{minipage}+
\begin{minipage}[bottom]{1in}
\includegraphics*[bb= 0 0 150 130,height=.75in]{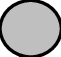}
\end{minipage}
\]
where, on the LHS we have a contribution which stems from a 2-d
model diagram with three propagators, however with two insertions of
the imaginary part of the propagator (i.e. $D_0$), and on the RHS we
have a sum of interacting diagrams of the 4-d theory, ${\cal N}=4$
SYM, and where all possible variants including scalar fields are
implied. The LHS contribution may be obtained precisely, as the
integrals over the points on the latitudes are evaluable. On the RHS
we find a by now well-known divergence cancellation between the last
two diagrams. We can then express everything in terms of finite
integrals over the bulk space-time interaction points. These in turn
can be analyzed in the limit where the two latitudes are coincident.
The results are that
\[
\begin{minipage}[bottom]{1in}
\includegraphics*[bb= 0 0 150 130,height=.75in]{X.eps}
\end{minipage}+
\begin{minipage}[bottom]{1in}
\includegraphics*[bb= 0 0 150 130,height=.75in]{H.eps}
\end{minipage}+
\begin{minipage}[bottom]{1in}
\includegraphics*[bb= 0 0 150 130,height=.75in]{IY.eps}
\end{minipage}+
\begin{minipage}[bottom]{1in}
\includegraphics*[bb= 0 0 150 130,height=.75in]{bubble.eps}
\end{minipage}<\quad {\cal O}(h^0)
\]
while,
\[
\begin{minipage}[bottom]{1in}
\includegraphics*[bb= 0 0 180 130,height=.75in]{3rungD.eps}
\end{minipage}\,~ \sim \quad |h|^2.
\]


\subsection{Preliminaries} \label{sec:prelim}

The latitudes we consider are given by

\be
W = \frac{1}{N} \Tr {\cal P} \exp \oint d\t \,
\left( i \,\dot x^\mu A_\mu + |\dot x| \Theta^I \,\Phi_I \right),
\ee

\ni where

\be
x^\m = (s\TO\cos\t,s\TO\sin\t,c\TO),\qquad \Theta^I = (-c\TO\cos\t,-c\TO\sin\t,s\TO),
\ee

\ni and where we have used the shorthand $c\TO \equiv \cos\TO$ and
similarly for $\sin$. The combined gauge field and scalar propagator
joining the two latitudes (in Feynman gauge) is then given by

\be\label{N4prop}
D_{12} \equiv
\frac{g^2}{4\pi^2}
\frac{-\dot x_1 \cdot \dot x_2 + |\dot x_1||\dot x_2| \Theta_1\cdot
  \Theta_2}
{(x_1-x_2)^2} = \frac{g^2}{4\pi^2}
\frac{s\TO^1s\TO^2\left[ \cos\t_{12}(c\TO^1 c\TO^2-1) +
    s\TO^1s\TO^2\right]}{2\left(1 - c\TO^1c\TO^2 - s\TO^1s\TO^2\cos\t_{12}\right)}.
\ee

\ni This ``loop-to-loop propagator'' is more compactly expressed as

\be
D_{12} = \frac{g^2}{4\pi^2} \frac{(1-c\TO^1c\TO^2)}{2}
\left(\frac{\cos\t_{12}+\L}{~~\cos\t_{12}+\L^{-1}}\right),\qquad
\L \equiv \frac{s\TO^1s\TO^2}{c\TO^1c\TO^2-1}.
\ee

\ni We are interested also in a reduced 2-d theory living on an
$S^2$ parametrized by the complex variable $z$ such that

\be
x^\m  = \frac{1}{1+z\bar z} ( z+ \bar z, -i(z- \bar z), 1-z\bar z)
\ee

\ni and so $z = e^{i\t} \tan(\TO/2)$ describes our latitudes. This
theory is pure gauge. Its fields are $A_z$ and $A_{\bar z}$. In the
light-cone gauge $A_{\bar z}=0$ while \cite{Drukker:2007yx,Drukker:2007qr}

\be
\la A_z(z)\,A_z(w) \ra = \frac{2g^2_{\text{2d}}}{\pi}
\frac{1}{(1+z\bar z)}\frac{1}{(1+w\bar w)} \frac{\bar z-\bar w}{z-w},
\ee

\ni where $g^2_{\text{2d}} = -g^2/(4\pi)$. In this theory we may also
construct the standard Wilson loop ${\tiny 1\over N }\Tr {\cal P} \exp i
\oint A\, dx$. The loop-to-loop propagator here is

\be
{\cal D}_{12}=
i^2\dot z_1 \dot z_2 \la A_z(z_1)\,A_z(z_2) \ra =
\frac{2g^2_{\text{2d}}}{\pi}\frac{s\TO^1 s\TO^2}{4}\left(
\frac{\l_1^2e^{-i\t_{12}} + \l_2^2e^{i\t_{12}} - 2 \l_1\l_2}
{\l_1^2+\l_2^2 - 2\l_1\l_2\cos\t_{12}}\right)
\ee

\ni where $\l_i = \tan(\TO^i/2)$. This can be put into a much more suggestive
form

\bsp\label{2dprop}
{\cal D}_{12} =
\frac{g^2}{4\pi^2} \frac{(1-c\TO^1c\TO^2)}{2}
\left(\frac{\cos\t_{12}+\L}{~~\cos\t_{12}+\L^{-1}}\right)
+i \frac{g^2}{4\pi^2} \frac{(c\TO^1-c\TO^2)}{2}
\left(\frac{\sin\t_{12}}{\cos\t_{12}+\L^{-1}}\right)
\end{split}
\ee

\ni where we see that the real component is exactly the loop-to-loop
propagator in the 4-d theory, i.e. $D_{12}$ defined in (\ref{N4prop}).


\subsection{A relation between diagrams} \label{sec:rel}

\begin{figure}
\begin{center}
\includegraphics*[bb= 0 0 150 130,height=0.75in]{2rung.eps}
\includegraphics*[bb= 0 0 150 130,height=0.75in]{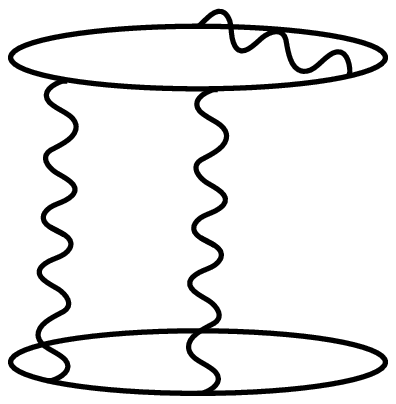}
\includegraphics*[bb= 0 0 150 130,height=0.75in]{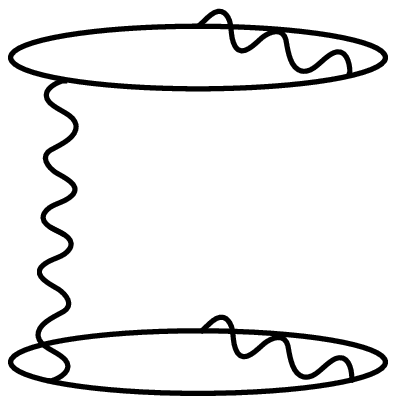}
\end{center}
\caption{These 2-d theory diagrams are trivially equivalent to their
  4-d counterparts.}
\label{fig:simrungs}
\end{figure}

We are interested in calculating the connected correlator between two
Wilson latitudes, both in the 2-d and 4-d theory. We begin with the
2-d calculation.  We perform calculations using the gauge group
$SU(N)$, in the large $N$ limit. Therefore we are interested only in
planar diagrams, while single insertions on a Wilson loop vanish since
the generators of $SU(N)$ are traceless. The 2-d theory, being in the
light-cone gauge, is free of interactions - it has only ladder
diagrams. In fact there are three 2-d ladder diagrams which are
trivially equivalent to those of the 4-d theory. These are pictured in
figure \ref{fig:simrungs}.
In the first two diagrams, due to the fact that at least one of the loops has
only two insertions, and due to the cyclicity of the trace, the
imaginary component of (\ref{2dprop}) integrates to zero since

\be
\int_0^{2\pi} d\T \frac{\sin\T}{\cos\T + \L^{-1}} = 0.
\ee

\ni Similarly, in the last diagram, any insertions of the imaginary
component of the loop-to-loop propagator vanish. Therefore only the
real component of the propagator contributes - giving precisely the
result for the 4-d theory. At order $\l^2$ the only non-vanishing
planar diagram in either theory is the two-rung ladder (pictured in
figure \ref{fig:simrungs} for the 2-d theory). Thus the two theories
agree at this level, however this is a direct result of the one-loop
proof given in \cite{Drukker:2007yx,Drukker:2007qr}.

Up to order $\l^3$ there is only one other planar ladder diagram - the
triple rung. The triple rung is given by

\begin{tabular}{ll}
\begin{minipage}{0.03\textwidth}
\hspace{-0.4in}
\vspace{+0.6in}
\includegraphics*[bb= 17 8 142 180,height=.65in]{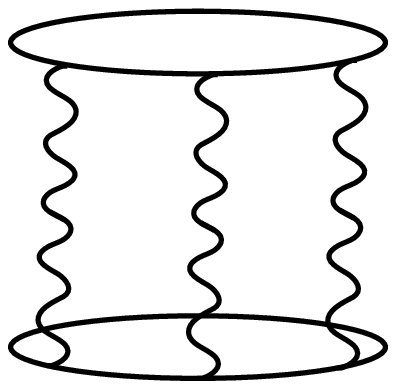}
\end{minipage}
&
\begin{minipage}{0.9\textwidth}
\bsp\nonumber
=\frac{N^3}{8N^2} \int_0^{2\pi}d\t_1  \int_0^{\t_1}d\t_2
\int_0^{\t_2}d\t_3
\int_0^{2\pi}d\s_1  \int_0^{\s_1}d\s_2
\int_0^{\s_2}d\s_3\,\Biggl\{ {\cal D}_{\s_1\,\t_1} \,{\cal D}_{\s_2\,\t_2}
\,{\cal D}_{\s_3\,\t_3} \\
+{\cal D}_{\s_1\,\t_3} \,{\cal D}_{\s_2\,\t_1}
\,{\cal D}_{\s_3\,\t_2}
+ {\cal D}_{\s_1\,\t_2} \,{\cal D}_{\s_2\,\t_3}
\,{\cal D}_{\s_3\,\t_1} \Biggr\}.
\end{split}
\ee
\end{minipage}
\end{tabular}\\

\ni Upon substitution of the 2-d theory propagator (\ref{2dprop}), we
see that the terms involving an odd number of insertions of the
imaginary component vanish, whereas clearly three insertions of the
real component gives exactly the triple rung in the 4-d theory. We are
therefore left with the following equality, should the 2-d theory
truly agree with the 4-d

\bsp\label{triprung}
\text{\Rmnum{3}}=&
\frac{\pi}{12N^2} \left(\frac{g^2N}{8\pi^2}\right)^3
\,i^2\,(c\TO^1-c\TO^2)^2(1-c\TO^1c\TO^2)
\int_0^{2\pi}d\phi \int_0^{2\pi}d\T_1
\int_0^{2\pi}d\T_2\int_0^{\T_1}d\psi_1 \int_0^{\T_2}d\psi_2\\
\Biggl\{ &\left(\frac{\cos(\phi+\T_2)+\L}{\cos(\phi+\T_2)+\L^{-1}}\right)\,
\left(\frac{\sin(\phi+\psi_2-\T_1)}{\cos(\phi+\psi_2-\T_1)+\L^{-1}}\right)\,
\left(\frac{\sin(\phi-\psi_1)}{\cos(\phi-\psi_1)+\L^{-1}}\right)\\
+&\left(\frac{\sin(\phi+\T_2)}{\cos(\phi+\T_2)+\L^{-1}}\right)\,
\left(\frac{\cos(\phi+\psi_2-\T_1)+\L}{\cos(\phi+\psi_2-\T_1)+\L^{-1}}\right)\,
\left(\frac{\sin(\phi-\psi_1)}{\cos(\phi-\psi_1)+\L^{-1}}\right)\\
+&\left(\frac{\sin(\phi+\T_2)}{\cos(\phi+\T_2)+\L^{-1}}\right)\,
\left(\frac{\sin(\phi+\psi_2-\T_1)}{\cos(\phi+\psi_2-\T_1)+\L^{-1}}\right)\,
\left(\frac{\cos(\phi-\psi_1)+\L}{\cos(\phi-\psi_1)+\L^{-1}}\right)\Biggr\}\\
&=\text{Sum of interacting diagrams of 4-d theory:
  X, H, IY, and 2-rung bubble}
\end{split}
\ee

\ni i.e., the triple-rung with two insertions of the imaginary
component of the loop-to-loop propagator ought to equal the sum of
all remaining diagrams of the 4-d theory - the so-called X, H, IY,
and 1-loop corrected two-rung ladder (or ``2-rung bubble'')
diagrams. We visit these diagrams individually in appendix
\ref{sec:diags}; they are depicted in figure \ref{fig:4ddiags}.

The integrations in (\ref{triprung}) can be carried out rather simply
because of the happy fact that

\be
\frac{\sin\phi}{\cos\phi + \L^{-1}} = -\p_\phi \ln \left(
  -\L^{-1}-\cos\phi \right)
\ee

\ni where we have ensured that the argument of the $\ln$ is always
positive. The result is\footnote{In an earlier version of this
  manuscript an error was present in (\ref{triprungint}) giving a
  mismatch with zero-instanton QCD$_2$.}

\bsp\label{triprungint}
\text{\Rmnum{3}}=&
\frac{\pi}{4N^2} \left(\frac{g^2N}{8\pi^2}\right)^3
\,i^2\,(c\TO^1-c\TO^2)^2 \Bigr(1-c\TO^1c\TO^2-|c\TO^1-c\TO^2|\Bigl)
(2\pi)^3\\
&\times \left[ -2\,\text{Li}_2\left(\frac{r^2-1}{r^2}\right)
 + 2 \ln r^2\ln\frac{r^2-1}{r^2} +
  \frac{\pi^2}{3} \right]
\end{split}
\ee

\ni where

\bsp
r \equiv \frac{1-c\TO^1 c\TO^2 + |c\TO^1 - c\TO^2|}{s\TO^1 s\TO^2},\quad
r^{-1} =  \frac{1-c\TO^1 c\TO^2 - |c\TO^1 - c\TO^2|}{s\TO^1 s\TO^2},\quad
\L^{-1} = -\frac{1}{2}(r+r^{-1}).
\end{split}
\ee

\ni We are therefore interested in whether or not this expression can
be recovered by the sum of interacting diagrams of the 4-d theory.

\subsection{Results}

We will investigate the proposed relation (\ref{triprung}) in the
limit in which the two latitudes are coincident. Looking at
(\ref{triprungint}) we see that in this limit (where $r\rightarrow
1$)

\be \text{\Rmnum{3}} \simeq
-\frac{\l^3}{N^2}\,\frac{s^2\TO}{2^8\cdot 3} \,(c\TO^1 -
c\TO^2)^2 \sim |h|^2. \ee

\ni The evaluation of the X, H, and IY diagrams are collected in
appendix \ref{sec:diags}. The results in the coincident limit $\TO^1
\simeq \TO^2 \simeq \TO$ are as follows

\[
\begin{minipage}[bottom]{1in}
\includegraphics*[bb= 0 0 150 130,height=.75in]{X.eps}
\end{minipage} = \frac{\l^3}{8N^2} \frac{1}{32} \,s^4\TO \,
|h|,\quad
\begin{minipage}[bottom]{1in}
\includegraphics*[bb= 0 0 150 130,height=.75in]{H.eps}
\end{minipage} 
<{\cal O}(h^0),
\]

\[
\begin{minipage}[bottom]{1in}
\includegraphics*[bb= 0 0 150 130,height=.75in]{IY.eps}
\end{minipage}+
\begin{minipage}[bottom]{1in}
\includegraphics*[bb= 0 0 150 130,height=.75in]{bubble.eps}
\end{minipage} 
<{\cal O}(|h|\log|h|).
\]

\section{Discussion}
\label{sec:end}

The stunning agreement found in section \ref{<w>} for the VEV of a
single Wilson loop at ${\cal O}(\l^2)$ is the result of an intriguing
cancellation of interacting Feynman diagrams with rainbow/ladders. It
certainly points to the capturing of these loops by a reduced model,
which for single Wilson loop VEV's agrees with the proposal made in
\cite{Drukker:2007qr}, while for Wilson loop correlators, does not
disagree. It remains a challenge to push the SYM correlator
calculation to further orders in $h$, or ideally, to an exact result
which could be matched against the 2-d model.

There are also further analyses which could be carried out. One of
these is to consider the connected correlator in the limit as one of
the latitudes shrinks to a point.  A similar limit was taken in the
work \cite{Arutyunov:2001hs}, for two 1/2 BPS circles. There it was
shown that the shrunken Wilson loop is given by a sum of local
operators, both protected and unprotected by supersymmetry. The
unprotected operators lead to terms which diverge as the logarithm of
the radius of the shrinking loop; these logarithms arise from the
interacting graphs and allow the determination of the operator's
anomalous dimension at first order in the 't Hooft coupling. It would
be interesting to repeat this analysis using the results collected
here; we leave this to a further publication. It would also be
interesting to compute the connected correlator at strong coupling,
using string theory; there two-point functions with protected
operators may be accessible
\cite{Berenstein:1998ij,Semenoff:2001xp,Giombi:2006de,Semenoff:2006am}.
If so, the summation of ladder diagrams along the lines of
\cite{Semenoff:2001xp,Semenoff:2006am} could be attempted in the gauge
theory and compared.


\section*{Acknowledgements}

It is a pleasure to thank Jan Plefka and Matthias Staudacher for
discussions, and Nadav Drukker for discussions and for suggesting these
calculations. The author would also like to thank the Galileo Galilei
Institute for Theoretical Physics for hospitality during the later
stages of completion of this work. This work was funded in part by a
Postdoctoral Fellowship from the Natural Sciences and Engineering
Research Council of Canada (NSERC), and also by the Volkswagen
Foundation.


\appendix
\section{Longitudes: divergence cancellation}
\label{sec:longitudes}

It is known that the two-loop diagrams with internal vertices
cancel-out for the 1/2 BPS circle. However, here, in the case of two
longitudes, we will not find the same cancellation. We find a finite
remainder, which is zero in the $\d=\pi$ limit. To begin, we re-cap
the cancellation mechanism for the 1/2 BPS circle. Equation (28) of
\cite{Erickson:2000af} gives the contribution from the triple vertex
diagram as

\begin{equation}
\begin{split}
\Sigma_3&=\l^2
\frac{\Gamma(2\omega-2)}{2^{2\omega+5}\pi^{2\omega}} \int_0^1
d\alpha\,d\beta\,d\gamma\,(\alpha\beta\gamma)^{\omega-2}\delta(1-\alpha-\beta-\gamma)
\\&
\qquad\times\oint d\tau_1\,d\tau_2\,d\tau_3\,
\frac{\epsilon(\tau_1\,\tau_2\,\tau_3) (1-\cos\tau_{13})
\bigl(\alpha(1-\alpha)\sin\tau_{12}+ \alpha\gamma\sin\tau_{23}\bigr)
} { \bigl[\alpha\beta(1-\cos\tau_{12})+
\beta\gamma(1-\cos\tau_{23})+ \gamma\alpha(1-\cos\tau_{13})
\bigr]^{2\omega-2} }.
\end{split}
\label{master}
\end{equation}

\ni By using the identity

\begin{equation}\label{ident}
\oint d\tau_1\,d\tau_2\,d\tau_3\,\frac{\partial}{\partial \tau_1}
\frac{\epsilon(\tau_1\,\tau_2\,\tau_3) (1-\cos\tau_{13}) } {
\Delta^{2\omega-3} } =0
\end{equation}

\ni where $ \Delta=\alpha\beta(1-\cos\tau_{12})+
\beta\gamma(1-\cos\tau_{23})+ \gamma\alpha(1-\cos\tau_{13})$, and
$\o=2$ on the physical dimension, one may relate $\S_3$ to the
one-loop-corrected, single-rung ladder diagram, and an extra piece
which vanishes on the physical dimension.

\subsection{Insertions on a single longitude}

\begin{figure}[ht]
\begin{center}
\includegraphics*[bb=22 70 290 260, height=1.in]{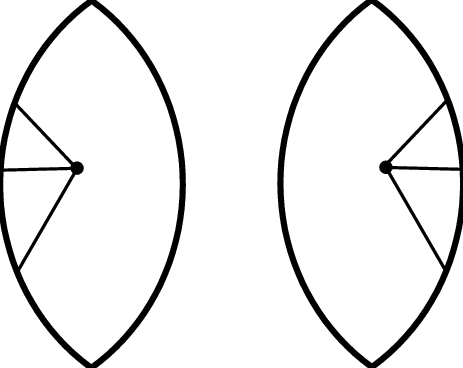}
\includegraphics*[bb=26 45 258 235, height=1.in]{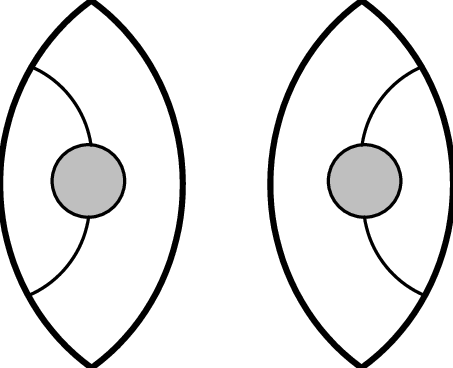}
\end{center}
\caption{Simplest class of triple vertex diagrams for the two
  longitudes. The solid lines refer to both scalars and gauge fields.}
\label{fig:triple}
\end{figure}

The simplest class of triple vertex diagrams for the two longitudes
are pictured in figure \ref{fig:triple}.

\ni We can use (\ref{master}) for these diagrams as well, the only
difference being the range of the loop parameters, which invalidates
(\ref{ident}). This means that after the cancellation of the
self-energy diagrams shown schematically in figure \ref{fig:triple},
there is a finite quantity left-over. If we take the range of the
$\t_i$ to be between $0$ and $\pi$, then we have that the RHS of
(\ref{ident}) is no longer zero but (under integration over
$\a,\b,\g$)

\bsp\label{02pi} \int_0^\pi d \t_2 \int_0^{\t_2} d\t_3 \, \Biggl\{
&\frac{\cos \t_3 - \cos \t_2}{\left[ \alpha\beta(1+\cos\tau_{2})+
\beta\gamma(1-\cos\tau_{23})+
\gamma\alpha(1+\cos\tau_{3})\right]^{2\o-3}}\\
&+ \frac{\cos \t_3 - \cos \t_2}{\left[ \alpha\beta(1-\cos\tau_{2})+
\beta\gamma(1-\cos\tau_{23})+
\gamma\alpha(1-\cos\tau_{3})\right]^{2\o-3}} \Biggr\}.
\end{split}
\ee

\ni The complement of this contribution, where the loop parameters
travel between $\pi$ and $2\pi$, is

\bsp\label{pi2pi} \int_\pi^{2\pi} d \t_2 \int_\pi^{\t_2} d\t_3 \,
\Biggl\{ &\frac{-\cos \t_3 + \cos \t_2}{\left[
\alpha\beta(1-\cos\tau_{2})+ \beta\gamma(1-\cos\tau_{23})+
\gamma\alpha(1-\cos\tau_{3})\right]^{2\o-3}}\\
&+ \frac{-\cos \t_3 + \cos \t_2}{\left[ \alpha\beta(1+\cos\tau_{2})+
\beta\gamma(1-\cos\tau_{23})+
\gamma\alpha(1+\cos\tau_{3})\right]^{2\o-3}} \Biggr\}.
\end{split}
\ee

\ni By shifting the loop parameters by $\pi$ in (\ref{pi2pi}) we
find that it is just equal to (\ref{02pi}). We will see that these
quantities are removed when we consider insertions between the two
longitudes.


\subsection{Insertions between the two longitudes}

\begin{figure}[ht]
\begin{center}
\includegraphics*[bb=22 73 490 262, height=1.in]{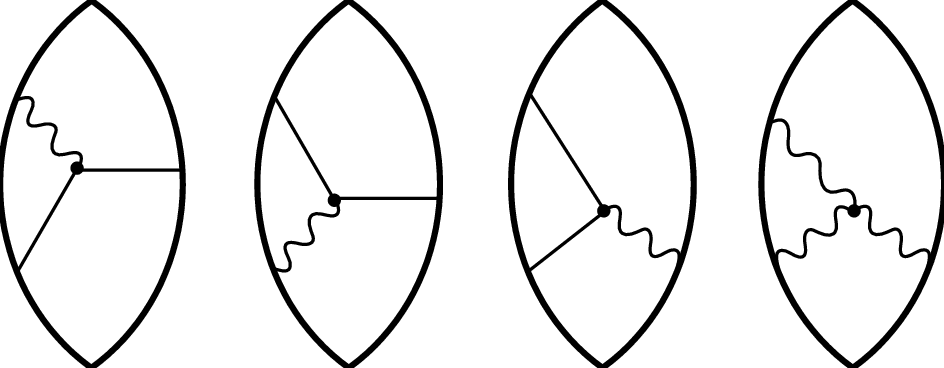}
\end{center}
\caption{Triple vertex diagrams which connect two longitudes. These
  diagrams do not cancel completely against the diagram shown in
  figure \ref{fig:corronerung}.}
\label{fig:triple2}
\end{figure}

The next class of triple vertex diagrams are those that connect the
two longitudes. In figure \ref{fig:triple2} we have shown those with
two insertions on the $0 \rightarrow \pi$ contour, however we must
equally consider those with two insertions on the opposite contour.
These diagrams can essentially be ``read-off'' from
(\ref{master}). The results are

\bsp\label{mysigma3} \S_3= \frac{\l^2}{4} \int_\pi^{2\pi} d\t_1
\int_0^\pi d\t_2 \int_0^{\pi} d\t_3 \, \e(\t_2 \,\t_3) \biggl\{
&\left( \dot y_1 \cdot \dot x_2 + \cos \d \right)
\dot x_3 \cdot \left(\p_{x_2} - \p_{y_1} \right) \\
+&\left( \dot x_2 \cdot \dot x_3 - 1 \right)
\dot y_1 \cdot \p_{x_3}  \biggr\} G(y_1, x_2, x_3)\\
-\frac{\l^2}{4} \int_\pi^{2\pi} d\t_1 \int_\pi^{2\pi} d\t_2
\int_0^{\pi} d\t_3\, \e(\t_1 \, \t_2) \biggl\{ &\left( \dot y_2
\cdot \dot x_3 + \cos \d \right)
\dot y_1 \cdot \left(\p_{y_2} - \p_{x_3} \right) \\
+&\left( \dot y_1 \cdot \dot y_2 - 1 \right) \dot x_3 \cdot \p_{y_1}
\biggr\} G(y_1, y_2, x_3)
\end{split}
\ee

\ni where

\bsp\nonumber &G(y_1,x_2,x_3) = \frac{\G(2\o-3)}{2^{2\o+3}
\pi^{2\o}} \int_0^1
d\alpha\,d\beta\,d\gamma\,(\alpha\beta\gamma)^{\omega-2}\delta(1-\alpha-\beta-\gamma)
\frac{1}{\D^{2\o-3} },\\
&G(y_1,y_2,x_3) = \frac{\G(2\o-3)}{2^{2\o+3} \pi^{2\o}} \int_0^1
d\alpha\,d\beta\,d\gamma\,(\alpha\beta\gamma)^{\omega-2}\delta(1-\alpha-\beta-\gamma)
\frac{1}{\wt\D^{2\o-3}
}
\end{split}
\ee

\ni where

\bsp \D = &\a\b ( 1+ \s \sin \t_1 \sin \t_2 - \cos \t_1 \cos
    \t_2 )\\
&+ \b \g (1 - \cos \t_{23}) + \a \g ( 1 + \s \sin \t_1 \sin \t_3 -
\cos \t_1 \cos \t_3 )\\
\wt\D = &\a \b (1 - \cos \t_{12}) +\b\g ( 1+ \s \sin \t_2 \sin \t_3
- \cos \t_2 \cos
    \t_3 )\\&+ \a \g ( 1 + \s \sin \t_1 \sin \t_3 -
\cos \t_1 \cos \t_3 )
\end{split}
\ee

\ni and where $\s \equiv \cos \d$. In fact the second expression is
just equal to the first, and so we are free to take twice the first
expression.

\begin{figure}[ht]
\begin{center}
\includegraphics*[bb=24 24 128 213, height=1.in]{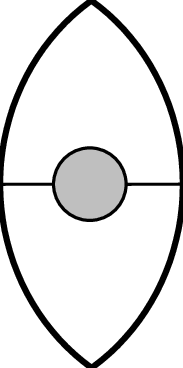}
\end{center}
\caption{The one-loop-corrected one rung ladder; it is divergent and
  must be cancelled by the diagrams shown in figure \ref{fig:triple2}.}
\label{fig:corronerung}
\end{figure}

Our strategy will be to generalize the mechanism used for the 1/2
BPS circle, described under equation (\ref{ident}), to the present
case. We will be looking to cancel out the divergent diagram shown
in figure \ref{fig:corronerung}.
This diagram gives the following contribution (see equation (12)
of \cite{Erickson:2000af})

\be\label{bubble} -\frac{\l^2 \G^2(\o-1)}{128 \pi^{2\o}
(2-\o)(2\o-3)} 2
 \int_{\pi}^{2\pi} dt_1 \int_0^{\pi} dt_2 \,
\frac{\cos\d \cos t_1 \cos t_2 - \sin t_1 \sin t_2 -
\cos\d}{\left[2(1 + \cos \d
  \sin t_1 \sin t_2 -\cos t_1 \cos t_2 )\right]^{2\o-3} }.
\ee

\ni Therefore we will consider

\bsp\label{derivative} &\p_{\t_3} \frac{(\s \cos \t_1 \cos \t_3 - \s
- \sin \t_1 \sin \t_3)}
{\D^{2\o-3}} \\
&=\frac{(2\o -3)}{\D^{2\o-2}} \Bigr[\s \cos \t_1 \cos \t_3 - \s -
\sin \t_1 \sin \t_3 \Bigl] \Bigl[ \b\g \sin\t_{23} - \a\g ( \s
\sin\t_1\cos\t_3 +
  \cos\t_1\sin\t_3) \Bigr]\\
&+ \frac{1}{\D^{2\o-2}} \Bigl[ \D \Bigr]\Bigl[ -\sin\t_1 \cos\t_3 -
\s \cos\t_1
  \sin\t_3 \Bigr]
\end{split}
\ee

\ni where $\D = \a\b ( 1+ \s \sin \t_1 \sin \t_2 - \cos \t_1 \cos
\t_2 ) + \b \g (1 - \cos \t_{23}) + \a \g ( 1 + \s \sin \t_1 \sin
\t_3 - \cos \t_1 \cos \t_3 )$. The first contribution from the
integrand in (\ref{mysigma3}) is

\bsp\label{integrand} A_1=\left( \dot y_1 \cdot \dot x_2 + \s
\right) \dot x_3 \cdot \p_{x_2} \frac{1}{\D^{2\o-3}} &= (3 - 2\o)
\Bigl[ \s( 1- \cos \t_1 \cos \t_2) + \sin \t_1 \sin \t_2 \Bigr]\\
&\times\Bigl[ \b(\a+\g) \sin \t_{23} + \a\b \left( \s \sin\t_1 \cos
\t_3 + \cos \t_1 \sin \t_3 \right) \Bigr] \frac{1}{\D^{2\o-2}}.
\end{split}
\ee

\ni We use (\ref{derivative}) to derive the following relation

\bsp\label{ident_b} & {\cal N} \int_\pi^{2\pi} d\t_1 \int_0^\pi
d\t_2 \int_0^{\pi}
d\t_3 \, \e(\t_2 \,\t_3) \, A_1 \\
&= {\cal N} (3-2\o) \Biggl[ \int_\pi^{2\pi} d\t_1 \int_0^\pi d\t_2
\frac{-\s-\s
  c_1}{\D^{2\o-3} |_{\t_3 = \pi}}
-  \int_\pi^{2\pi} d\t_1 \int_0^\pi d\t_2 \frac{\s-\s
  c_1}{\D^{2\o-3} |_{\t_3 = 0}}\Biggr]\\
&~ +{\cal N} (3-2\o) \int_\pi^{2\pi} d\t_1 \int_0^\pi d\t_2
\int_0^{\pi}
d\t_3 \, \e(\t_2 \,\t_3) \, \frac{B_1}{\D^{2\o-2}}\\
&-2\l^2\frac{\G(2\o-3)}{2^{2\o+3} \pi^{2\o}}
  \frac{\G^2(\o-1)\,\G(2-\o)}{\G(2\o-2)\,\G(5-\o)}
 \int_\pi^{2\pi} d\t_1 \int_0^\pi
d\t_2 \frac{ \s-\s c_1 c_2 + s_1 s_2}{[(1+\s s_1 s_1 - c_1
  c_2)]^{2\o-3}}
\end{split}
\ee

\ni where $c_i \equiv \cos \t_i$, $s_i \equiv \sin \t_i$, finite terms
multiplied by $(2\o-4)$ have been suppressed, $B_1$ is given along
with similar contributions from the other portions of the integrand in
(\ref{mysigma3}) in (\ref{Bees}), and we have introduced the notation

\be {\cal N} \equiv \frac{\l^2}{2} \frac{\G(2\o-3)}{2^{2\o+3}
\pi^{2\o}} \int_0^1 d\alpha\,d\beta\,d\gamma\,
\delta(1-\alpha-\beta-\gamma) (\alpha\beta\gamma)^{\omega-2}. \ee

\ni In the limit $\o\rightarrow 2$ the singular contribution in the
last line of (\ref{ident_b}) cancels (\ref{bubble}) exactly. The first
two terms represent finite quantities left over from this
cancellation. We may now continue and derive similar identities for
the remaining terms in (\ref{mysigma3}). Continuing with the second
term in the first integral of (\ref{mysigma3})

\bsp A_2 =-\left( \dot y_1 \cdot \dot x_2 + \s \right) \dot x_3
\cdot \p_{y_1} &\frac{1}{\D^{2\o-3}} = (3 - 2\o)
\Bigl[ \s( 1- \cos \t_1 \cos \t_2) + \sin \t_1 \sin \t_2 \Bigr]\\
&\times\Bigl[ \a\b \sin \t_{23} + \a(\b+\g) \left( \s \sin\t_1 \cos
\t_3 + \cos \t_1 \sin \t_3 \right) \Bigr] \frac{1}{\D^{2\o-2}}
\end{split}
\ee

\ni we use the derivative

\bsp\label{derivative2} -\p_{\t_3} \frac{(1-\cos\t_{23})}
{\D^{2\o-3}} =&-\frac{(2\o -3)}{\D^{2\o-2}} \Bigr[1-\cos\t_{23}
\Bigl] \Bigl[ \b\g \sin\t_{23} - \a\g ( \s \sin\t_1\cos\t_3 +
  \cos\t_1\sin\t_3) \Bigr]\\
&+ \frac{1}{\D^{2\o-2}}\Bigl[ \D \Bigr] \Bigl[ \sin\t_{23} \Bigr]
\end{split}
\ee

\ni to derive 

\bsp\label{A2} {\cal N} \int_\pi^{2\pi} &d\t_1 \int_0^\pi d\t_2
\int_0^{\pi}
d\t_3 \, \e(\t_2 \,\t_3) \, A_2 \\
&=
 {\cal N} (3-2\o) \int_\pi^{2\pi} d\t_1 \int_0^\pi d\t_2 \int_0^{\pi}
d\t_3 \, \e(\t_2 \,\t_3) \, \frac{B_2}{\D^{2\o-2}} \\
&+ {\cal N} (3-2\o) \Biggl[ \int_\pi^{2\pi} d\t_1 \int_0^\pi d\t_2
\frac{-(1+
  c_2)}{\D^{2\o-3} |_{\t_3 = \pi}}
-  \int_\pi^{2\pi} d\t_1 \int_0^\pi d\t_2 \frac{1-
  c_2}{\D^{2\o-3} |_{\t_3 = 0}}\Biggr].
\end{split}
\ee

\ni Similarly for the third term in (\ref{mysigma3}), we have

\bsp A_3 &= \left( \dot x_2 \cdot \dot x_3 - 1\right) \dot y_1 \cdot
\p_{x_3} \frac{1}{\D^{2\o-3}} = (3 - 2\o)
\Bigl[ \cos\t_{23} - 1 \Bigr]\\
&\times\Bigl[ \b\g \left( \s \cos \t_1 \sin \t_2 + \sin \t_1 \cos
\t_2 \right) - \g(\a+\b) \left( \s \cos\t_1 \sin \t_3 +
\sin \t_1 \cos \t_3 \right) \Bigr] \frac{1}{\D^{2\o-2}}\\
\end{split}
\ee

\ni and we use the derivative

\bsp\label{derivative3} &-\p_{\t_1} \frac{(1-\cos\t_{13})}
{\D^{2\o-3}} \\
&=\frac{(2\o -3)}{\D^{2\o-2}} \Bigr[1-\cos\t_{13}\Bigl] \Bigl[ \a\b
(\s\cos\t_1\sin\t_2+\sin\t_1\cos\t_2)+ \a\g ( \s \cos\t_1\sin\t_3 +
  \sin\t_1\cos\t_3) \Bigr]\\
&+ \frac{1}{\D^{2\o-2}} \Bigl[ \D \Bigr]\Bigl[-\sin\t_{13}\Bigr].
\end{split}
\ee

\ni We find

\bsp {\cal N} \int_\pi^{2\pi} &d\t_1 \int_0^\pi d\t_2 \int_0^{\pi}
d\t_3 \, \e(\t_2 \,\t_3) \, A_3 \\
&=
 {\cal N} (3-2\o) \int_\pi^{2\pi} d\t_1 \int_0^\pi d\t_2 \int_0^{\pi}
d\t_3 \, \e(\t_2 \,\t_3) \, \frac{B_3}{\D^{2\o-2}} \\
&+ {\cal N} (3-2\o) \Biggl[ \int_0^{\pi} d\t_2 \int_0^\pi d\t_3
\,\e(\t_2\,\t_3)\,
 \frac{(1-c_3)}{\D^{2\o-3} |_{\t_3 = 2\pi}}
-  \int_0^{\pi} d\t_2 \int_0^\pi d\t_3\,\e(\t_2\,\t_3)\, \frac{(1+
c_3)}{\D^{2\o-3} |_{\t_3 = \pi}}\Biggr].
\end{split}
\ee

\ni The last line above cancels the contributions of (\ref{02pi})
and (\ref{pi2pi}) exactly. We are now in a position to quote the
finite result of the internal vertex diagrams, it is given by

\bsp \L_3 = -\frac{\hat \l^2}{16} \int_0^1 d\a\, d\b\, d\g\,
\d(1-\a-\b-\g) \Biggl[  \int_\pi^{2\pi} d\t_1 \int_0^\pi d\t_2
\int_0^{\pi}
d\t_3 \, \e(\t_2 \,\t_3) \, \frac{B_1+B_2+B_3}{\D^2}\\
- \int_\pi^{2\pi} d\t_1 \int_0^\pi d\t_2
\frac{(1+\s)(2+c_1+c_2)}{[\a\b(1+\s s_1 s_2 - c_1 c_2) + \b\g(1+c_2)
+ \a\g(1+c_1)]} \Biggr]
\end{split}
\ee

\ni where we have combined the surface terms from (\ref{ident_b})
and (\ref{A2}). A simple expression for the sum of $B_1$, $B_2$, and
$B_3$ is given by

\bsp \label{Bees}
B_1+B_2+B_3 = &\a\g (\s^2-1) \left[ 2 s_1 (c_3 - c_2) - s_1 c_1
(1-\cos \t_{23}) \right]\\
&+ \a\g(\s+1)(s_2-s_3)(c_3-c_1)\\
&+\a\g(\s+1)\left[\sin\t_{13}^+ - \sin \t_{12}^++\sin\t_{23} \right]\\
&+\a\g(\s+1) \sin\t_{23}(1-\cos\t_{13}^+) +\b\g (\s+1) c_1 s_3
(1-\cos\t_{23})
\end{split}
\ee

\ni where we have introduced some shorthand $\t_{ij}\equiv \t_i -
\t_j$, $\t_{ij}^+ \equiv \t_i + \t_j$. It is clear that at $\delta =
\pi$, where $\s=-1$, $\L_3$ is explicitly zero, as it must be, in
order to coincide with the known results of the 1/2 BPS circle.


\section{Connected correlator: interacting diagrams}
\label{sec:diags}

\begin{figure}
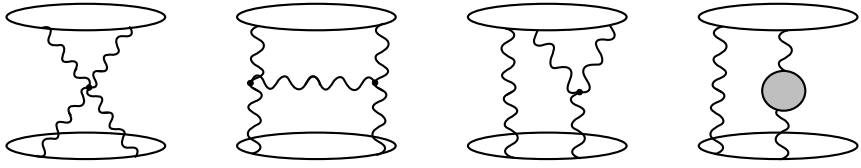

\begin{center}
\includegraphics*[bb= 0 0 150 130,height=1.0in]{X.eps}
\includegraphics*[bb= 0 0 150 130,height=1.0in]{H.eps}
\includegraphics*[bb= 0 0 150 130,height=1.0in]{IY.eps}
\includegraphics*[bb= 0 0 150 130,height=1.0in]{bubble.eps}
\end{center}
\caption{The interacting diagrams of the 4-d theory which contribute
  to the planar, connected correlator of two Wilson loops. The gauge
  field lines are understood to also represent scalars - as allowed by
the couplings of ${\cal N}=4$ SYM.} \label{fig:4ddiags}
\end{figure}

In this section we undertake the calculation of the diagrams
depicted\footnote{There is also a second IY diagram, where the two
  latitudes are exchanged.} in figure \ref{fig:4ddiags}, in the 4-d
theory, i.e. ${\cal N}=4$ SYM.
We employ the Euclidean action in Feynman gauge, the details of
which (along with the conventions used here) are to be found in
\cite{Plefka:2001bu} and \cite{Erickson:2000af}. We will find that
there is a divergence cancellation between the IY and 2-rung bubble,
completely analogous to the one found for the case of two 1/2 BPS
Wilson circles in \cite{Plefka:2001bu}. The X and H diagrams will
also yield extremely similar - but not exactly the same - results as
those found in \cite{Plefka:2001bu} for the 1/2 BPS case. Due to the
great similarity between the following calculation and that
performed in \cite{Plefka:2001bu}, we will not be overly explicit.
The reader is referred to \cite{Plefka:2001bu} for further details.

The general strategy is to perform the integrals over the Wilson
loop contours themselves, leaving the bulk integrations over the
space-time points of interaction unevaluated. We will use the
notation $x(\t)$ and $y(\s)$ to refer to the parametrizations of the
two Wilson loops at polar angles $\TO^1$ and $\TO^2$ respectively,
along with their associated scalar paths $\Theta_x(\t)$ and
$\Theta_y(\s)$. We will find the following integrals repeatedly
useful

\bsp\label{eyes} &I(\TO) \equiv \int_0^{2\pi}
d\t\,\frac{1}{a+b\cos\t+c\sin\t} =
\frac{2\pi}{\sqrt{a^2-(b^2+c^2)}},\\
&I_c(\TO) \equiv \int_0^{2\pi} d\t\,\frac{\cos\t}{a+b\cos\t+c\sin\t}
= \frac{2\pi\,b\,\left(\sqrt{a^2-(b^2+c^2)} -
    a\right)}{\sqrt{a^2-(b^2+c^2)}\,(b^2+c^2)},\\
&I_s(\TO) \equiv \int_0^{2\pi} d\t\,\frac{\sin\t}{a+b\cos\t+c\sin\t}
= \frac{2\pi\,c\,\left(\sqrt{a^2-(b^2+c^2)} -
    a\right)}{\sqrt{a^2-(b^2+c^2)}\,(b^2+c^2)},
\end{split}
\ee

\ni where

\be\label{abc} a = \r^2 + w_3^2 + s^2\TO +(w_2-c\TO)^2,\qquad b=
-2\, s\TO \,w_0,\qquad c= -2\, s\TO \,w_1, \ee

\ni and $\r^2 = w_0^2 + w_1^2$, where $w=(w_0,w_1,w_2,w_3)$ is a
space-time interaction point. We will also make use of some further
shorthand

\bsp\label{Rs} R_1\equiv \r^2 + w_2^2 +w_3^2,\quad &R_2
\equiv (\r+\cot\TO)^2 + (w_2 -1)^2 + w_3^2 .
\end{split}
\ee

\subsection{X diagram}

The X-diagram is given by

\bsp X=\frac{8 g^6 N^3}{4^3 N^2}\int_0^{2\pi}d\t_1
\int_0^{2\pi}d\t_2&
\int_0^{2\pi}d\s_1 \int_0^{2\pi}d\s_2\\
 & \Biggl[ \Bigl( \dot x_1 \cdot
  \dot y_2 - \Theta_{x_1} \cdot \Theta_{y_2} |\dot x_1||\dot y_2|
  \Bigr) \Bigl( \dot x_2 \cdot \dot y_1 - \Theta_{x_2} \cdot
  \Theta_{y_1} |\dot x_2||\dot y_1| \Bigr)\\ -&\Bigl( \dot x_1 \cdot
  \dot x_2 - \Theta_{x_1} \cdot \Theta_{x_2} |\dot x_1||\dot x_2|
  \Bigr) \Bigl( \dot y_1 \cdot \dot y_2 - \Theta_{y_1} \cdot
  \Theta_{y_2} |\dot y_1||\dot y_2| \Bigr)\Biggr]\\ &\times
\left(\frac{1}{4\pi^2}\right)^4 \int\frac{d^4
  w}{(x_1-w)^2(x_2-w)^2(y_1-w)^2(y_2-w)^2}\\
\end{split}
\ee

\ni Evaluating the integrals over $\t_1,\t_2,\s_1,\s_2$, we have

\bsp\label{X} X= \frac{8 g^6 N^3}{4^3
N^2}&\left(\frac{1}{4\pi^2}\right)^4 \int d^4 w\, \Biggl\{ s^2\TO^1
s^2\TO^2 (1-c\TO^1 c\TO^2)^2 \Bigl[ I_c(\TO^1) I_c(\TO^2) +
  I_s(\TO^1) I_s(\TO^2) \Bigr]^2\\
&-2s^3\TO^1 s^3\TO^2 (1-c\TO^1 c\TO^2) I(\TO^1) I(\TO^2) \Bigl[
I_c(\TO^1) I_c(\TO^2) + I_s(\TO^1) I_s(\TO^2) \Bigr]\\
&+s^4\TO^1 s^4\TO^2 \Bigl( I^2(\TO^2) \bigl[I_c^2(\TO^1) +
I_s^2(\TO^1)\bigl] +  I^2(\TO^1) \bigl[I_c^2(\TO^2) +
I_s^2(\TO^2)\bigl] \Bigl)\\
&-s^4\TO^1 s^4\TO^2 \Bigl( I_c^2(\TO^1) + I_s^2(\TO^1)\Bigl)
\Bigl(I_c^2(\TO^2) + I_s^2(\TO^2) \Bigr) \Biggl\}
\end{split}
\ee

\ni where the $I,I_c,I_s$ are given by (\ref{eyes}), (\ref{abc}).


\subsection{H diagram}

The H diagram is most compactly expressed in terms of an extended
notation

\be \dot x^M \equiv (\dot x_\m; -i|\dot x| \Theta^I),\qquad \p_{x^M}
\equiv (\p_{x_\m};0) \ee

\ni with $\m = 0,\ldots,3$ and $I=4,\ldots,9$, so that

\be \dot x^M = s\TO \,(-s\t,\,c\t,\,0,\,0;\,i\,c\TO\, c\t,\,
i\,c\TO\, s\t,\, -i\, s\TO,\,0,\,0,\,0). \ee

\ni The contribution of this diagram is given by

\be\label{Hfund} H = \frac{\l^3}{8N^2}
\left(\frac{1}{4\pi^2}\right)^5 \int d^4 w \int d^4z\,\frac{H^M(w)
\, H^M(z)}{(w-z)^2} \ee

\ni where

\bsp H^M(w) \equiv \int_0^{2\pi} d\t \int_0^{2\pi}d\s \, \Biggl[
 2 \dot y^M \bigl(\dot x \cdot \p_{y}\bigr)
-2 \dot x^M \bigl(\dot y \cdot \p_{x}\bigr) +\Bigl[ \dot x \cdot
\dot y - \Theta_{x} \cdot \Theta_{y}
  |\dot x||\dot y| \Bigr]\bigl( \p_{x^M} - \p_{y^M} \bigr)
  \Biggr] \\\times\frac{1}{(x-w)^2(y-w)^2}
\end{split}
\ee

\ni and $x=x(\t)=(s\TO^1 c\t,s\TO^1 s\t,c\TO^1)$, $y=y(\s)=(s\TO^2
c\s,s\TO^2 s\s,c\TO^2)$, etc. One finds

\bsp\label{HI} &H^4(w) = -2  i\,s\TO^1s\TO^2(c\TO^1-c\TO^2)  \Bigl(
I_s(\TO^2) \left(\p_{w_0}  I_c(\TO^1) \right) -
I_c(\TO^2) \left( \p_{w_0}  I_s(\TO^1) \right) \Bigr)\\
&H^5(w) = -2  i\,s\TO^1s\TO^2(c\TO^1-c\TO^2)  \Bigl( I_s(\TO^2)
\left( \p_{w_1} I_c(\TO^1) \right) -
I_c(\TO^2) \left( \p_{w_1}  I_s(\TO^1)\right) \Bigr)\\
&H^6(w) = 0
\end{split}
\ee

\bsp\label{Hmu} H^\m(w) = s\TO^1 s\TO^2 (1-c\TO^1 c\TO^2)&\Bigl[
I_c(\TO^1) \left( \p_{w_\m} I_c(\TO^2) \right) - I_c(\TO^2) \left(
\p_{w_\m} I_c(\TO^1) \right)\\ +& I_s(\TO^1) \left( \p_{w_\m}
I_s(\TO^2) \right) -
I_s(\TO^2) \left( \p_{w_\m} I_s(\TO^1) \right) \Bigr]\\
-s^2\TO^1 s^2\TO^2 &\Bigl[ I(\TO^1) \left( \p_{w_\m} I(\TO^2)
\right) - I(\TO^2) \left( \p_{w_\m} I(\TO^1) \right)\Bigr].
\end{split}
\ee


\subsection{IY and two-rung bubble divergence cancellation}

In this subsection we will demonstrate the cancellation of the
divergence stemming from the two-rung bubble against the divergent
part of the IY diagram. The finite parts left-over from this
cancellation are calculated. The strategy follows
\cite{Plefka:2001bu} closely; Feynman parameters are introduced in
favour of bulk integrations in order to demonstrate the
cancellation, then the finite left-overs are re-cast in terms of
bulk integrations.

The IY diagram is given by

\bsp\label{IY} IY = \frac{\l^3}{8N^2} \int_0^{2\pi} d\vt\, {\cal
F}(\vt) \oint d\t_1\,d\t_2\,d\s_1\,E(\t_1\,\t_2)\, \Biggl\{
D(\t_1,\s_1) \left[ \dot x_2 \cdot \p_{y_1} - \dot x_2 \cdot
    \p_{x_1} \right]\\
 + D(\t_1,\t_2)\, \dot y_1 \cdot \p_{x_1} \Biggr\}
G(x_1,x_2,y_1)
\end{split}
\ee

\ni where

\be {\cal F}(\vt) = -\frac{(1-c \TO^1 c \TO^2)}{8\pi^2}
\,\frac{\cos\vt +
  \L}{\cos\vt + \frac{1}{\L}},\qquad \L = \frac{s \TO^1 s \TO^2}{c
  \TO^1 c\TO^2-1},
\ee

\ni and

\be D(\t, \s) = s\TO^1 s\TO^2 \left[(1-c\TO^1 c\TO^2)\cos(\t-\s)
 -s\TO^1 s\TO^2\right], \qquad
D(\t_1,\t_2) = s^4\TO^1 \,(\cos\t_{12}  -1), \ee

\ni and

\be E(\t_1 \, \t_2) \equiv 2\pi\, \text{sgn}(\t_1 - \t_2) - 2\,(\t_1
- \t_2). \ee

\ni The triple-vertex kernel $G(x_1,x_2,y_1)$ is given in
dimensional regularization ($d=2\o$) by

\be G(x_1,x_2,y_1) = \frac{\G(2\o-3)}{2^6 \pi^{2\o}} \int_0^1 d\a\,
d\b\, d\g\, \frac{(\a\b\g)^{\o-2}\d(1-\a-\b-\g)}{[\a\b(x_1-x_2)^2 +
\b\g(x_2-y_1)^2 + \a\g(x_1 -y_1)^2]^{2\o-3}}. \ee

\ni We rewrite (\ref{IY}) as

\bsp\label{IYno2} IY = \frac{\l^3}{8N^2} \int_0^{2\pi} d\vt\, {\cal
F}(\vt)& \oint
d\t_1\,d\t_2\,d\s_1\,\frac{\G(2\o-3)}{2^6 \pi^{2\o}} \\
&\times  2(2\o-3)\,\int_0^1 d\a\, d\b\, d\g\,
(\a\b\g)^{\o-2}\d(1-\a-\b-\g)\, {\cal O},
\end{split}
\ee

\ni where

\bsp\label{to} {\cal O} \equiv \frac{E(\t_1\,\t_2)}{\D^{2\o-2}}
\biggl\{&s\TO^1 s\TO^2 \Bigl(\cos(\t_1-\s_1)[c\TO^1 c\TO^2-1]+s\TO^1
\TO^2\Bigr)\Bigl( (2\a+\b)\g\, s\TO^1 s\TO^2 \sin(\s_1-\t_2)\\ &-
(2\g + \b) \a\, s^2\TO^1 \sin\t_{12} \Bigr) +s^4\TO^1
(1-\cos\t_{12})\, (2\b+\g)\a\, s\TO^1 s\TO^2 \sin(\t_1-\s_1)
\biggr\},
\end{split}
\ee

\ni and where

\bsp \D = 2\a\b \,s^2\TO^1\, (1-\cos\t_{12}) + 2\b\g\,\Bigl(1-s\TO^1
s\TO^2 \cos(\t_2-\s_1) -c\TO^1 c\TO^2\Bigr)\\ +
2\a\g\,\Bigl(1-s\TO^1 s\TO^2 \cos(\t_1-\s_1) -c\TO^1 c\TO^2\Bigr).
\end{split}
\ee

\ni After \cite{Plefka:2001bu} we consider the following total
derivatives

\bsp K_1 = -\frac{1}{2}\,(\cos\TO^1 \cos\TO^2 - 1)\, \sin^2\TO^1\,
\p_{\t_2} &\left( E(\t_1\,\t_2)
  \frac{(1-\cos\t_{12})}{\D^{2\o-3}} \right)
=\\& -\frac{s^2\TO^1}{2}\,(c\TO^1 c\TO^2 - 1)\, \left[
-4\pi\d(\t_{12}) +2 \right]
\frac{(1-\cos\t_{12})}{\D^{2\o-3}}\\
&+ \frac{s^2\TO^1}{2} \,(c\TO^1 c\TO^2 - 1)\,
E(\t_1\,\t_2)\,\frac{\sin\t_{12}}{\D^{2\o-3}}\\
&-s^2\TO^1\,(c\TO^1 c\TO^2 - 1)\,E(\t_1\,\t_2)\,(3-2\o)(1-\cos\t_{12})\\
&\qquad\times\frac{-\a\b s^2\TO^1\sin\t_{12}-\b\g s\TO^1 s\TO^2
\sin(\s_1-\t_2)}{\D^{2\o-2}}
\end{split}
\ee

\bsp K_2 =
 s \TO^1 s \TO^2 \,\p_{\t_1} &\left( E(\t_1\,\t_2)
  \frac{\cos(\t_1-\s_1)[ c \TO^1 c \TO^2-1]
 + s \TO^1 s \TO^2}{\D^{2\o-3}} \right) =\\
&s\TO^1 s\TO^2 \left[ 4\pi \d(\t_{12}) -2\right]
\frac{\cos(\t_1-\s_1)[ c \TO^1 c \TO^2-1]
 + s \TO^1 s \TO^2}{\D^{2\o-3}}\\
&- s\TO^1 s \TO^2 E(\t_1\,\t_2) [c\TO^1 c\TO^2-1]
\frac{\sin(\t_1-\s_1)}{\D^{2\o-3}}\\
&+s \TO^1 s \TO^2 \,2\,(3-2\o)\,E(\t_1\,\t_2) \left(\cos(\t_1-\s_1)[
c \TO^1 c \TO^2-1]
 + s \TO^1 s \TO^2\right)\\
&\qquad\times\frac{ \left[ \a\b s^2\TO^1 \sin\t_{12} - \a\g s \TO^1
s \TO^2 \sin(\s_1 - \t_1) \right]}{\D^{2\o-2}}
\end{split}
\ee

\bsp K_3 = \frac{ s \TO^1 s \TO^2}{2} \,\p_{\t_2} &\left(
E(\t_1\,\t_2)
  \frac{\cos(\t_1-\s_1)[ c \TO^1 c \TO^2-1]
 + s \TO^1 s \TO^2}{\D^{2\o-3}} \right)=\\
& \frac{s\TO^1 s\TO^2}{2} \left[ -4\pi \d(\t_{12}) +2\right]
\frac{\cos(\t_1-\s_1)[ c \TO^1 c \TO^2-1]
 + s \TO^1 s \TO^2}{\D^{2\o-3}}\\
&+s \TO^1 s \TO^2 \,(3-2\o)\,E(\t_1\,\t_2) \left(\cos(\t_1-\s_1)[ c
\TO^1 c \TO^2-1]
 + s \TO^1 s \TO^2\right)\\ &\qquad \times\frac{ \left[
-\a\b s^2\TO^1 \sin\t_{12} - \b\g s \TO^1 s \TO^2 \sin(\s_1 - \t_2)
\right]}{\D^{2\o-2}}
\end{split}
\ee

\ni The sum of the three RHS's may be expressed as follows (where we
use manipulations valid under the integrations in (\ref{IYno2}))

\bsp\label{match}
K_1+&K_2 +K_3 = \\
&2\pi \d(\t_{12})\,s \TO^1 s \TO^2 \, \frac{\cos(\t_1-\s_1)[ c \TO^1
c \TO^2-1]
 + s \TO^1 s \TO^2}{\D^{2\o-3}}\\
&- s^2\TO^1\,(c\TO^1 c\TO^2 - 1)\,
\frac{(1-\cos\t_{12})}{\D^{2\o-3}} -s \TO^1 s \TO^2
\frac{\cos(\t_1-\s_1)[ c \TO^1 c \TO^2-1]
 + s \TO^1 s \TO^2}{\D^{2\o-3}}\\
&+\frac{s^2 \TO^1}{2} \,(c\TO^1 c\TO^2 - 1)\,E(\t_1\,\t_2)
\frac{\sin\t_{12}}{\D^{2\o-3}} + s \TO^1 s\TO^2 \, E(\t_1\,\t_2)
(1-c\TO^1 c\TO^2)
\frac{\sin(\t_1-\s_1)}{\D^{2\o-3}}\\
&+(3-2\o) \,s^2\TO^1 \,(c\TO^1 c\TO^2 - 1)\,E(\t_1\,\t_2)\,
(1-\cos\t_{12}) \frac{s^2\TO^1 \a \b \sin\t_{12} + s\TO^1 s\TO^2
\b\g \sin(\s_1-\t_2)}
{\D^{2\o-2}}\\
&+(3-2\o) \,s\TO^1 s \TO^2 \, E(\t_1\,\t_2) \,\left(\cos(\t_1-\s_1)[
c \TO^1 c \TO^2-1]
 + s \TO^1 s \TO^2\right) \\
&\times\frac{s^2\TO^1 \a \b \sin\t_{12} -2 s\TO^1
 s\TO^2 \a\g \sin(\s_1-\t_1) -s\TO^1 s\TO^2 \b\g \sin(\s_1-\t_2) }{\D^{2\o-2}}.
\end{split}
\ee

\ni We would now like to reconstitute (\ref{to}) using the terms
proportional to $E(\t_1\,\t_2)$ in (\ref{match}). We do this by
first stripping-off terms proportional to $(4-2\o)$ by writing
$(3-2\o) = (4-2\o) -1$. We define

\bsp \Psi \equiv \frac{E(\t_1\,\t_2)}{\D^{2\o-2}} \biggl\{
&\frac{s^2 \TO^1}{2}\,(c\TO^1 c\TO^2 - 1)\,  \sin\t_{12}\, \D + s
\TO^1 s\TO^2 \,  (1-c\TO^1 c\TO^2)
\sin(\t_1-\s_1)\,\D\\
 - &s^2\TO^1 \,(c\TO^1 c\TO^2 - 1)\, (1-\cos\t_{12})
\Bigl( s^2\TO^1 \a \b \sin\t_{12} + s\TO^1 s\TO^2 \b\g
\sin(\s_1-\t_2)
\Bigr)\\
- &s\TO^1 s \TO^2 \,\left(\cos(\t_1-\s_1)[ c \TO^1 c \TO^2-1]
 + s \TO^1 s \TO^2\right) \\
&\qquad\times\Bigr(s^2\TO^1 \a \b \sin\t_{12} -2 s\TO^1
 s\TO^2 \a\g \sin(\s_1-\t_1) -s\TO^1 s\TO^2 \b\g \sin(\s_1-\t_2)
 \Bigl) \biggr\},
\end{split}
\ee

\ni so that

\bsp \sum_i&K_i = \Psi + 2\pi \d(\t_{12})\,s \TO^1 s \TO^2 \,
\frac{\cos(\t_1-\s_1)[ c \TO^1 c \TO^2-1]
 + s \TO^1 s \TO^2}{\D^{2\o-3}}\\
&- s^2\TO^1\,(c\TO^1 c\TO^2 - 1)\,
\frac{(1-\cos\t_{12})}{\D^{2\o-3}} -s \TO^1 s \TO^2
\frac{\cos(\t_1-\s_1)[ c \TO^1 c \TO^2-1]
 + s \TO^1 s \TO^2}{\D^{2\o-3}}\\
&+(4-2\o) \,s^2\TO^1 \,(c\TO^1 c\TO^2 - 1)\,E(\t_1\,\t_2)\,
(1-\cos\t_{12}) \frac{s^2\TO^1 \a \b \sin\t_{12} + s\TO^1 s\TO^2
\b\g \sin(\s_1-\t_2)}
{\D^{2\o-2}}\\
&+(4-2\o) \,s\TO^1 s \TO^2 \, E(\t_1\,\t_2) \,\left(\cos(\t_1-\s_1)[
c \TO^1 c \TO^2-1]
 + s \TO^1 s \TO^2\right) \\
&\qquad\times\frac{s^2\TO^1 \a \b \sin\t_{12} -2 s\TO^1
 s\TO^2 \a\g \sin(\s_1-\t_1) -s\TO^1 s\TO^2 \b\g \sin(\s_1-\t_2)
}{\D^{2\o-2}},
\end{split}
\ee

\ni then, expressing the last two terms with derivatives, we have

\bsp
 = &\Psi + 2\pi \d(\t_{12})\,s \TO^1 s \TO^2 \,
\frac{\cos(\t_1-\s_1)[ c \TO^1 c \TO^2-1]
 + s \TO^1 s \TO^2}{\D^{2\o-3}}\\
&- s^2\TO^1\,(c\TO^1 c\TO^2 - 1)\,
\frac{(1-\cos\t_{12})}{\D^{2\o-3}} -s \TO^1 s \TO^2
\frac{\cos(\t_1-\s_1)[ c \TO^1 c \TO^2-1]
 + s \TO^1 s \TO^2}{\D^{2\o-3}}\\
&+\frac{(4-2\o)}{(3-2\o)} \,\frac{s^2\TO^1}{2} \,(c\TO^1 c\TO^2 -
1)\,E(\t_1\,\t_2)\, (1-\cos\t_{12}) \left(-\p_{\t_2}\right)\frac{1}
{\D^{2\o-3}}\\
&+\frac{(4-2\o)}{(3-2\o)} \,s\TO^1 s \TO^2 \, E(\t_1\,\t_2) \,
\left(\cos(\t_1-\s_1)[ c \TO^1 c \TO^2-1]
 + s \TO^1 s
 \TO^2\right)\left(\p_{\t_1}+\frac{1}{2}\p_{\t_2}\right)\frac{1}{\D^{2\o-3}},
\end{split}
\ee

\ni then using integration by parts in $\t_1,\t_2$,

\bsp = &\Psi -\frac{ 2\pi \d(\t_{12})}{(3-2\o)}\,s \TO^1 s \TO^2 \,
\frac{\cos(\t_1-\s_1)[ c \TO^1 c \TO^2-1]
 + s \TO^1 s \TO^2}{\D^{2\o-3}}\\
&+\frac{ s^2\TO^1\,(c\TO^1 c\TO^2 - 1)}{(3-2\o)}\,
\frac{(1-\cos\t_{12})}{\D^{2\o-3}} +\frac{s \TO^1 s \TO^2}{(3-2\o)}
\frac{\cos(\t_1-\s_1)[ c \TO^1 c \TO^2-1]
 + s \TO^1 s \TO^2}{\D^{2\o-3}}\\
&-\frac{(4-2\o)}{(3-2\o)}\,\frac{E(\t_1\,\t_2)}{\D^{2\o-3}}\,(c\TO^1
c\TO^2 - 1) \, \Bigl( \frac{s^2\TO^1}{2} \, \sin\t_{12} - s\TO^1
s\TO^2 \sin(\t_1-\s_1) \Bigr).
\end{split}
\ee

\ni One can then show that

\bsp {\cal O} = \Psi + \frac{E(\t_1\,\t_2)}{\D^{2\o-2}}\, (c\TO^1 -
c\TO^2)^2\,\g(1-\g)\,\Bigl( s^2\TO^1 \sin \t_{12} - 2 s\TO^1
s \TO^2 \sin(\t_1-\s_1) \Bigr)\\
+ \frac{E(\t_1\,\t_2)}{\D^{2\o-2}}\,s^3\TO^1 s\TO^2 c\TO^1 (c \TO^2
- c\TO^1) (2\b+\g)\a (1-\cos\t_{12})\,\sin(\t_1-\s_1),
\end{split}
\ee

\ni and therefore

\be {\cal O} = \text{total deriv.} + (IY)_{SE} + (IY)_1 + (IY)_2 +
(IY)_3 + (IY)_{\o-2}, \ee

\ni since $\sum K_i$ is a total derivative, and where we have
introduced

\bsp\label{iy1se} &(IY)_{SE} = \frac{ 2\pi \d(\t_{12})}{(3-2\o)}\,s
\TO^1 s \TO^2 \, \frac{\cos(\t_1-\s_1)[ c \TO^1 c \TO^2-1]
 + s \TO^1 s \TO^2}{\D^{2\o-3}}\\
&(IY)_1 = -\frac{ s^2\TO^1\,(c\TO^1 c\TO^2 - 1)}{(3-2\o)}\,
\frac{(1-\cos\t_{12})}{\D^{2\o-3}} -\frac{s \TO^1 s \TO^2}{(3-2\o)}
\frac{\cos(\t_1-\s_1)[ c \TO^1 c \TO^2-1]
 + s \TO^1 s \TO^2}{\D^{2\o-3}}
\end{split}
\ee \bsp\label{iy23} &(IY)_2 = \frac{E(\t_1\,\t_2)}{\D^{2\o-2}}\,
(c\TO^1 - c\TO^2)^2\,\g(1-\g)\,\Bigl( s^2\TO^1 \sin \t_{12} - 2
s\TO^1
s \TO^2 \sin(\t_1-\s_1) \Bigr)\\
&(IY)_3 = \frac{E(\t_1\,\t_2)}{\D^{2\o-2}}\,s^3\TO^1 s\TO^2 c\TO^1
(c \TO^2 -
c\TO^1) (2\b+\g)\a (1-\cos\t_{12})\,\sin(\t_1-\s_1)\\
&(IY)_{\o-2} =
\frac{(4-2\o)}{(3-2\o)}\,\frac{E(\t_1\,\t_2)}{\D^{2\o-3}} \,(c\TO^1
c\TO^2 - 1) \, \Bigl( \frac{s^2\TO^1}{2} \, \sin\t_{12} - s\TO^1
s\TO^2 \sin(\t_1-\s_1) \Bigr).
\end{split}
\ee

\ni Plugging these forms back into (\ref{IYno2}) one finds that half
of the 1-loop corrected two-rung diagram is canceled by $(IY)_{SE}$
(the $\TO^1 \lr \TO^2$ piece takes care of the other half), and that
$(IY)_{\o-2}$ is zero on the physical dimension $\o=2$. The
remaining terms $(IY)_{1,2,3}$ are finite on the physical dimension
and must be evaluated (along with their $\TO^1 \lr \TO^2$
counterparts). In the following subsections we recast $(IY)_{1,2,3}$
in terms of bulk integrations.


\subsubsection{$(IY)_1$}

Plugging $(IY)_1$ from (\ref{iy1se}) into (\ref{IYno2}), and
reverting to bulk integration, one finds

\bsp \Pi_1= &\frac{\l^3}{4N^2} \frac{1}{4\pi} \left(c\TO^1 c\TO^2 -
1 +|c\TO^1-c\TO^2| \right)\frac{1}{64\pi^6}
\oint d\t_1\, d\t_2\, d\s_1 \\
&\times\int d^4w\,\frac{
 s^2\TO^1\,(c\TO^1 c\TO^2 - 1)\,(1-\cos\t_{12})
+s \TO^1 s \TO^2 \Bigl(\cos(\t_1-\s_1)[ c \TO^1 c \TO^2-1]
 + s \TO^1 s \TO^2\Bigr)}{(x_1-w)^2\,(x_2-w)^2\,(y_1-w)^2},
\end{split}
\ee

\ni where we have used the result

\be \int_0^{2\pi} d\vt \,{\cal F}(\vt) = \frac{1}{4\pi} \left(c\TO^1
c\TO^2 - 1 +|c\TO^1-c\TO^2| \right). \ee

\ni We now continue by integrating over $\t_1,\t_2$, and $\s_1$. We
find

\bsp\label{pione} \Pi_1 = \frac{\l^3}{4N^2} &\frac{1}{4\pi}
\left(c\TO^1 c\TO^2 - 1 +|c\TO^1-c\TO^2| \right)\frac{1}{64\pi^6}\int d^4w\\
&\times\Bigl[
s^2\TO^1\,c\TO^2\,(c\TO^1-c\TO^2)\,I^2(\TO^1)\,I(\TO^2) -
s^2\TO^1\,(c\TO^1 c\TO^2-1) \left( I^2_c(\TO^1) +
  I^2_s(\TO^1)\right)\,I(\TO^2)\\
&\qquad+ s\TO^1 s\TO^2 \,(c\TO^1 c\TO^2-1)
\,\left(I_c(\TO^1)\,I_c(\TO^2) +
  I_s(\TO^1)\,I_s(\TO^2) \right)\,I(\TO^1) \Bigr].
\end{split}
\ee


\subsubsection{$(IY)_2$}

This contribution is significantly more complicated due to the
presence of the $E(\t_1\,\t_2)$ in the integrand. Plugging $(IY)_2$
from (\ref{iy23}) into (\ref{IYno2}), we need to evaluate

\bsp \Pi_2 = \frac{\l^3}{4N^2} &\frac{1}{4\pi} \left(c\TO^1 c\TO^2 -
1 +|c\TO^1-c\TO^2| \right) (c\TO^1 -
c\TO^2)^2\,\frac{1}{2\,(c\TO^1-c\TO^2)}\frac{\p}{\p(c\TO^2)}
\frac{1}{64\pi^6}\int d^4w\\
&\times  \oint d\t_1\, d\t_2\, d\s_1 \,E(\t_1\,\t_2) \frac{\Bigl(
s^2\TO^1 \sin \t_{12} - 2 s\TO^1 s \TO^2 \sin(\t_1-\s_1)
\Bigr)}{(x_1-w)^2\,(x_2-w)^2\,(y_1-w)^2},
\end{split}
\ee

\ni where we treat $c\TO^2$ and $s\TO^2$ as independent variables
for the purposes of differentiation, and therefore must be cautious
not to use trigonometric identities which relate them until after
the derivative has been taken. With this prescription

\bsp\label{del} \D = 2\a\b s^2\TO^1 (1-\cos\t_{12}) &+ \b\g \left[
s^2\TO^1 + s^2 \TO^2 + (c\TO^1 - c\TO^2)^2 -2 s\TO^1
  s\TO^2 \cos(\t_2-\s_1) \right]  \\
&+\a\g \left[ s^2\TO^1 + s^2 \TO^2 + (c\TO^1 - c\TO^2)^2 -2 s\TO^1
  s\TO^2 \cos(\t_1-\s_1) \right],
\end{split}
\ee

\ni and hence the factor $\g(1-\g)$ in (\ref{iy23}) is obtained
through the derivative in $c\TO$. The evaluation of the integrals
over $\t_1,\t_2$, and $\s_1$ are as in \cite{Plefka:2001bu}. The
results are

\bsp\label{C1} {\cal C}_1 \equiv \oint d\t_1\, d\t_2\, &d\s_1
\,E(\t_1\,\t_2)\frac{ \sin
  \t_{12}
}{(x_1-w)^2\,(x_2-w)^2\,(y_1-w)^2}\\
& = \frac{64
\pi^3}{\sqrt{a_1^2-(b_1^2+c_1^2)}\sqrt{a_2^2-(b_2^2+c_2^2)}}
\,\frac{a_1}{(b_1^2+c_1^2)} \ln
\left(\frac{a_1+\sqrt{a_1^2-(b_1^2+c_1^2)}}{2\sqrt{a_1^2-(b_1^2+c_1^2)}}
\right)
\end{split}
\ee

\bsp\label{C2} {\cal C}_2 \equiv \oint d\t_1\, d\t_2\, d\s_1
\,&E(\t_1\,\t_2)\frac{ \sin(
  \t_{1}-\s_1)
}{(x_1-w)^2\,(x_2-w)^2\,(y_1-w)^2}\\
& = \frac{32
\pi^3}{\sqrt{a_1^2-(b_1^2+c_1^2)}\sqrt{a_2^2-(b_2^2+c_2^2)}}
\,\frac{b_1 b_2 + c_1
  c_2}{(b_1^2+c_1^2)\left[a_2+\sqrt{a_2^2-(b_2^2+c_2^2)}\right]}\\
&\qquad\qquad\qquad\qquad\qquad\qquad\times \ln
\left(\frac{a_1+\sqrt{a_1^2-(b_1^2+c_1^2)}}{2\sqrt{a_1^2-(b_1^2+c_1^2)}}
\right)
\end{split}
\ee

\ni where the $\{a,b,c\}_i$ are given by (\ref{abc}) and where the
index refers to either $\TO^1$ or $\TO^2$. We therefore have that

\bsp\label{pitwo} \Pi_2 = \frac{\l^3}{4N^2} &\frac{1}{4\pi}
\left(c\TO^1 c\TO^2 - 1 +|c\TO^1-c\TO^2| \right)
\frac{(c\TO^1-c\TO^2)}{2} \frac{1}{64\pi^6}\int d^4w \,\p_{c\TO^2}
\left( s^2\TO^1\, {\cal C}_1 -
  2s\TO^1s\TO^2\,{\cal C}_2 \right).
\end{split}
\ee


\subsubsection{$(IY)_3$}

Looking at (\ref{iy23}) we see that we must express the integrand

\be \frac{E(\t_1\,\t_2)}{\D^{2\o-2}} ( 2\a\b+\a\g) (1-\cos\t_{12})
  \sin (\t_1 - \s_1)
\ee

\ni without Feynman parameters. Referring to (\ref{del}), and again
treating $s\TO$ and $c\TO$ as independent, we see that

\bsp &\p_{s\TO^1} \D = 4\a\b s\TO^1(1-\cos\t_{12}) + 2\b\g (s\TO^1 -
s\TO^2\cos(\t_2-\s_1)) + 2\a\g (s\TO^1 - s\TO^2\cos(\t_1-\s_1))\\
&\p_{s\TO^2} \D = 2\b\g (s\TO^2 - s\TO^1\cos(\t_2-\s_1)) + 2\a\g
(s\TO^2 - s\TO^1\cos(\t_1-\s_1))
\end{split}
\ee

\ni and therefore

\be \left( s\TO^1\p_{s\TO^1} - s\TO^2\p_{s\TO^2} \right) \D =  4\a\b
s^2\TO^1(1-\cos\t_{12}) + 2\g(1-\g) \left(s^2\TO^1-s^2\TO^2\right).
\ee

\ni Whereas

\bsp &(1-\cos\t_{12})\p_{\t_1} \D = (1-\cos\t_{12}) \left[ 2\a\b
s^2\TO^1 \sin\t_{12} + 2\a\g s\TO^1 s\TO^2 \sin(\t_1-\s_1)
\right]\\
&\qquad = \sin\t_{12} \Bigl[ \D - \b\g \left[ s^2\TO^1 + s^2 \TO^2 +
(c\TO^1 - c\TO^2)^2 -2 s\TO^1
  s\TO^2 \cos(\t_2-\s_1) \right]  \\
&\qquad\qquad-\a\g \left[ s^2\TO^1 + s^2 \TO^2 + (c\TO^1 - c\TO^2)^2
-2 s\TO^1
  s\TO^2 \cos(\t_1-\s_1) \right] \Bigr]\\
&\qquad\qquad+ 2\a\g s\TO^1 s\TO^2 \sin(\t_1-\s_1) (1-\cos\t_{12})\\
&\qquad =\sin\t_{12} \Bigl[ \D-\g(1-\g) \left[ s^2\TO^1-s^2\TO^2 +
  (c\TO^1-c\TO^2)^2\right] -s\TO^2\p_{s\TO^2} \D \Bigr]\\
&\qquad\qquad+ 2\a\g s\TO^1 s\TO^2 \sin(\t_1-\s_1) (1-\cos\t_{12}).
\end{split}
\ee

\ni We therefore have that

\bsp \frac{ ( 2\a\b+\a\g) (1-\cos\t_{12})
  \sin (\t_1 - \s_1)}{\D^{2\o-2}} = \frac{ \left( s\TO^1\p_{s\TO^1} - s\TO^2\p_{s\TO^2}\right)
  }{2(3-2\o)s^2\TO^1}\frac{\sin(\t_1-\s_1)}{\D^{2\o-3}}  \\
-\g(1-\g)\frac{\left(s^2\TO^1-s^2\TO^2\right)}{s^2\TO^1}\frac{\sin(\t_1-\s_1)}{\D^{2\o-2}}
+\frac{(1-\cos\t_{12})}{2(3-2\o)s\TO^1s\TO^2}
\p_{\t_1}\frac{1}{\D^{2\o-3}}\\
-\frac{\sin\t_{12}}{2 s\TO^1 s\TO^2} \frac{1}{\D^{2\o-3}}
+\frac{\sin\t_{12}}{2(3-2\o)s\TO^1}\,\p_{s\TO^2}\frac{1}{\D^{2\o-3}}\\
+\g(1-\g)\sin\t_{12}\frac{\left[ s^2\TO^1-s^2\TO^2 +
  (c\TO^1-c\TO^2)^2\right]}{2s\TO^1s\TO^2}\frac{1}{\D^{2\o-2}}.
\end{split}
\ee

\ni We now can express the $\g(1-\g)$ as a derivative in $c\TO^2$,
as per the previous subsection. Further we note that under
integration (and for $\o=2$)

\be
E(\t_1\,\t_2)\,\Biggl[\frac{(1-\cos\t_{12})}{2(3-2\o)s\TO^1s\TO^2}
\p_{\t_1}\frac{1}{\D^{2\o-3}} -\frac{\sin\t_{12}}{2 s\TO^1 s\TO^2}
\frac{1}{\D^{2\o-3}}\Biggr] =
\frac{(1-\cos\t_{12})}{(3-2\o)s\TO^1s\TO^2} \frac{1}{\D^{2\o-3}} \ee

\ni by integration by parts in $\t_1$. The RHS is then integrated as
per $(IY)_1$, and expressed in terms of $I(\TO),I_c(\TO),I_s(\TO)$.
The $(IY)_3$ contribution from (\ref{iy23}), once plugged-in to
(\ref{IYno2}), is then expressed as

\bsp\label{pithree} \Pi_3 = \frac{\l^3}{4N^2} &\frac{1}{4\pi}
\left(c\TO^1 c\TO^2 - 1 +|c\TO^1-c\TO^2| \right) \,s^3\TO^1\, s\TO^2
\,c\TO^1 (c \TO^1 -c\TO^2)
\frac{1}{64\pi^6}\int d^4w\\
&\times \Biggl\{ \frac{1}{2 s\TO^1s\TO^2}\left( s\TO^2 \p_{s\TO^2} +
c\TO^2 \p_{c\TO^2}
\right)\, {\cal C}_1\\
&+\frac{1}{2s^2\TO^1} \left(
  s\TO^1\p_{s\TO^1}-s\TO^2\p_{s\TO^2}
-\left(c\TO^1+c\TO^2\right)\p_{c\TO^2}\right)\,{\cal C}_2\\
&+\frac{1}{s\TO^1 s\TO^2}\,I(\TO^2) \left[ I^2(\TO^1) -
I_c^2(\TO^1)-
  I_s^2(\TO^1) \right] \Biggl\}.
\end{split}
\ee

\subsection{The coincident limit}

In what follows we will make extensive use of the integrals collected
in (\ref{eyes}). To simplify things we cast them in a simpler form

\bsp
I(\TO) = \frac{2\pi}{\sqrt{L^+ L^-}},\quad
I_c(\TO) = \frac{4\pi s\TO w_0}{\sqrt{L^+ L^-} (a + \sqrt{L^+ L^-})},\quad
I_s(\TO) = \frac{4\pi s\TO w_1}{\sqrt{L^+ L^-} (a + \sqrt{L^+ L^-})},\\
\text{where}\quad L^{\pm} \equiv (\r \pm s\TO)^2 +(w_2-c\TO)^2+w_3^2,\quad
a \equiv \r^2 + s^2\TO +(w_2-c\TO)^2+w_3^2.
\end{split}
\ee

\ni The $L^\pm$ and $a$ are found also in the ${\cal C}_i$
integrals defined in (\ref{C1}) and (\ref{C2}).
In order to extract
the leading behaviour as $\TO^1 \rightarrow \TO^2$ we can borrow from
the analysis carried out in \cite{Plefka:2001bu} for the case of
1/2 BPS circles in the coincident limit. The observation is that the
integration over the bulk space-time interaction point $w$ is dominated
in this limit by the region $\r = \sqrt{w_0^2 + w_1^2} \simeq s\TO^1\simeq
s\TO^2$, $w_2 \simeq c\TO^1 \simeq c\TO^2$, and $w_3 \simeq 0$. This is the
region where $L^- \simeq 0$, $L^+ \simeq 4s^2\TO$, and $a \simeq 2
s^2\TO$. In the special case of the H diagram where there are two
space-time interaction points, both integrations are dominated by this
same region. The $\TO^1 \rightarrow \TO^2$ limit is then taken as
follows:

\begin{enumerate}
\item Combine integrands with their $\TO^1 \lr \TO^2$ counterparts.
\item In the denominator, replace $L^+$ by $4s^2\TO$, $a$ by $2s^2\TO$,
  and $a+\sqrt{L^+ L^-}$ by $2s^2\TO$.
\item Shift $\r \rightarrow \r + s\TO^1$, $w_2 \rightarrow w_2 +
  c\TO^1$.
\item Scale all components of $w$ by $|h|$, where $h \equiv c\TO^1-c\TO^2$.
\item If resulting integral is divergent as $w\rightarrow \infty$, it
  should be cut-off at ${\cal O}(1/h)$.
\end{enumerate}

\subsubsection{X diagram}

The leading result for the X-diagram goes as $|h|$, specifically one
finds from (\ref{X})

\bsp X &\simeq \frac{\l^3}{8N^2} \left(\frac{1}{4\pi^2}\right)^4
8 \pi^8 s^4\TO \, |c\TO^1-c\TO^2|\\
& =  \frac{\l^3}{8N^2} \frac{1}{32} \,s^4\TO \, |h|.
\end{split}
\ee

\subsubsection{H diagram}

\ni Taking the coincident limit, we find that the leading terms are of
order $|h|^0$ and come from $H^\m$, i.e. (\ref{Hmu}). These are then
cancelled by their $\TO^1 \lr \TO^2$ counterparts. We give an
example of such a term below, which stems from the first two
components of $H^\m$
\bsp
&\frac{\l^3}{N^2} \frac{s^5\TO}{2^{12}\pi^4} |h| \int_{-\infty}^\infty
d\r\,d\bar\r\,dw_2\,dw_3\,dz_2\,dz_3\,\\
&\frac{\r+\cot\TO}{\sqrt{R_1(w)}R_2(w)} \frac{1}{\sqrt{(\r -\bar \r)^2
    +(w_2-z_2)^2 +(w_3-z_3)^2}}
\frac{\bar \r+\cot\TO}{\sqrt{R_1(z)}R_2(z)} ={\cal O}(h^0)
\end{split}
\ee
where the $R_i$ are defined in (\ref{Rs}). The integral is linearly
divergent and therefore supplies a factor of $1/|h|$. The $\TO^1 \lr
\TO^2$ counterpart cancels this contribution. We find that all such
contributions from $H^\m$ behave as above, and we therefore have that
\be
H < {\cal O}(h^0).
\ee

\subsubsection{IY diagram}

\ni In taking the coincident limit one finds that the leading contributions
go as $|h|\log|h|$ and come from $(IY)_3$, i.e. (\ref{pithree}). One finds the contribution
\bsp
&-\frac{\l^3}{N^2} \frac{s^2\TO c\TO}{256 \pi^3}
|h|\log|h| \int_{-\infty}^\infty d\r\,d w_2\, d w_3\\
&\times \frac{2\r(\r^2+w_2^2+w_3^2) + \cot\TO(3\r^2+w_2^2+w_3^2) -2w_2 \r +
  \r(1+\cot^2\TO)}{R_1(w)^{3/2} R_2(w)^{3/2}} = 0
\end{split}
\ee
where the $R_i$ are defined in (\ref{Rs}). The integral happens to
evaluate to zero, and is in any case cancelled by the $\TO^1 \lr
\TO^2$ counterpart. Therefore we have that
\be
(IY)_1+(IY)_2+(IY)_3 < {\cal O}(|h|\log|h|).
\ee


\end{document}